\documentclass{article}

\usepackage{arxiv}

\usepackage[utf8]{inputenc} 
\usepackage[T1]{fontenc}    
\usepackage{hyperref}       
\usepackage{url}            
\usepackage{booktabs}       
\usepackage{amsfonts}       
\usepackage{nicefrac}       
\usepackage{microtype}      
\usepackage{lipsum}		
\usepackage{graphicx}
\usepackage{xcolor}
\usepackage{dsfont}
\usepackage{amsmath}
\usepackage{amssymb}

\title{Finslerian wormholes in squared trace gravity}

\author{ \href{https://orcid.org/0000-0003-3126-2321}{\includegraphics[scale=0.06]{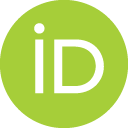}\hspace{1mm}Z. Nekouee} \\
	School of Physics, Damghan University\\ Damghan, 3671641167, Iran \\
	\texttt{zohrehnekouee@gmail.com} \\
	\And
	\href{https://orcid.org/0000-0002-8239-5741}{\includegraphics[scale=0.06]{orcid.png}\hspace{1mm}B. R. Yashwanth} \\
	Department of Mathematics, Mangalore Institute of Technology \& Engineering\\ Badaga Mijar, Moodabidri, Mangalore-574225, Karnataka, India \\
	\texttt{yashmath0123@gmail.com} \\
\And
	\href{https://orcid.org/0000-0002-8239-5741}{\includegraphics[scale=0.06]{orcid.png}\hspace{1mm}Manjunath Malligawad} \\
	Department of Mathematics, Sri Siddhartha Institute of Technology\\ A Constituent College of Siddhartha Academy of Higher
Education\\ Tumkur, 572105, Karnataka, India \\
	\texttt{manjunathmalligawad91@gmail.com} \\
\And
	\href{https://orcid.org/0000-0002-2711-4636}{\includegraphics[scale=0.06]{orcid.png}\hspace{1mm}S. K. Narasimhamurthy} \\
	Department of PG Studies and Research in Mathematics, Kuvempu University\\ Jnana Sahyadri, Shankaraghatta-577 451, Shivamogga, Karnataka, India \\
	\texttt{nmurthysk@gmail.com} \\
\And
	\href{https://orcid.org/0000-0003-0951-414X}{\includegraphics[scale=0.06]{orcid.png}\hspace{1mm}S. K. J. Pacif} \\
	Pacif Institute of Cosmology and Selfology (PICS)\\ Sagara, Sambalpur, 768 224, Odisha, India\\
Research Center of Astrophysics and Cosmology, Khazar University\\ Baku, AZ1096, 41 Mehseti Street, Azerbaijan\\
	\texttt{shibesh.math@gmail.com} \\
\And
	\href{https://orcid.org/0000-0001-9720-8817}{\includegraphics[scale=0.06]{orcid.png}\hspace{1mm}Kazuharu Bamba} \\
	Faculty of Symbiotic Systems Science, Fukushima University\\ Fukushima 960-1296, Japan\\
	\texttt{bamba@sss.fukushima-u.ac.jp} \\
}

\hypersetup{
pdftitle={Finslerian wormholes in squared trace gravity},
pdfsubject={q-bio.NC, q-bio.QM},
pdfauthor={Z. Nekouee et al.},
pdfkeywords={First keyword, Second keyword, More},
}

\begin{document}
\maketitle

\begin{abstract}
	We investigate traversable wormholes in squared-trace extended gravity within the framework of Finsler-Randers geometry equipped with the Barthel connection. The Einstein-Hilbert action is modified by terms involving the trace of the energy-momentum tensor and its square, generating effective anisotropies through matter-curvature coupling. The resulting field equations are studied under barotropic equations of state with exponential and power-law shape functions. Finslerian anisotropy introduces novel pressure dynamics that enable the classical energy conditions to be satisfied in specific parameter domains. Our analysis quantitatively maps how the Barthel connection scales the parameter space for non-exotic, geometrically viable wormholes compared to purely Riemannian models. These findings suggest that Finslerian modifications provide a powerful mechanism for realizing consistent wormhole structures, offering new perspectives on anisotropic and geometrically enriched space-time configurations in extended gravity.
\end{abstract}

\keywords{Wormhole \and Squared-trace extended gravity \and Non-exotic matter \and Finsler-Randers geometry \and Barthel connection}

\section{Introduction}\label{sec1}
The idea of traversable wormholes, idealized channels between separate regions of space-time, has intrigued physicists since their theoretical construction in general relativity (GR) \cite{Morris, Visser1995}. Within the standard GR context, these wormholes usually require the use of exotic matter that breaks the null energy condition (NEC) to stabilize geometry \cite{Hochberg1997, Visser1989,Katsuragawa}. Nevertheless, exotic matter is hypothetical, and its inconsistency with established physics has driven the study of alternative theories of gravity to create wormholes without contradicting the energy conditions \cite{Lobo2017}. Among all the modified theories~\cite{padmanabhan2007,Durrer2008,Sotiriou:2008rp,DeFelice2010,Nojiri:2010wj,Capozziello2011,Clifton:2011jh,Nojiri:2017ncd,De Falco,Capozziello}, $\mathfrak{f}(\mathds{R},\mathds{T})$ gravity has received significant interest, where $\mathds{R}$ is the Ricci scalar and $\mathds{T}$ represents the trace of the energy-momentum tensor.
\par This requirement severely limits the physical plausibility of the geometries. Alternative theories of gravity provide a natural means of relaxing this constraint by altering the effective gravitational dynamics, thereby allowing the support of wormhole structures with minimal or no exotic matter. One auspicious approach arises within the $\mathfrak{f}(\mathds{R},\mathds{T})$ class of theories \cite{Harko2011,Fortunato,Shukla,Wazny,Lamaaoune}, in which the gravitational Lagrangian depends not only on the Ricci scalar, but also on the trace of the energy--momentum tensor \cite{Moraes2018, Zubair2016, Rosa2022}. This matter-geometry coupling introduces extra degrees of freedom that can modify the effective energy-momentum tensor seen by the geometry, allowing the energy conditions to be satisfied under suitable circumstances \cite{Cipriano}.
A notable extension of this framework is the so-called \emph{squared trace-extended gravity} \cite{Board:2017ign, Nari:2018aqs, Akarsu:2018zxl}, characterized by $\mathfrak{f}(\mathds{R},\mathds{T}) = \mathds{R} + \ell_1 \mathds{T} + \ell_2 \mathds{T}^2,$ where $\ell_1$ and $\ell_2$ are coupling constants controlling the strength of the linear and quadratic trace contributions. The presence of the quadratic $\mathds{T}^2$ term represents higher-order matter corrections to the gravitational action. This addition plays a crucial role: it offers enhanced flexibility in tailoring the effective stress-energy tensor, thereby expanding the parameter space in which NEC, as well as other classical energy conditions (weak, dominant, and strong), can be satisfied while maintaining the flaring-out condition at the wormhole throat \cite{Parsaei, Sharif2023}.
\par Within this setting, a variety of shape functions, such as power-law or exponential profiles, can be employed along with different matter configurations, including anisotropic fluids and barotropic equations of state of the form $p_r = \omega_r \rho$ and $p_t = \omega_t \rho$ \cite{Tripathy}. Model's viability depends sensitively on the choice of coupling constants, matter distribution, and equation of state (EoS) parameters. For suitable choices, asymptotically flat wormhole geometries have been realized that remain entirely supported by non-exotic matter throughout the space-time.
\par From a broader perspective, the $\mathds{R} + \ell_1 \mathds{T} + \ell_2 \mathds{T}^2$ framework stands out among alternative theories of gravity for wormhole physics because it achieves this balance without resorting to elaborate curvature invariants or higher-derivative terms in $\mathds{R}$. Its appeal lies in its minimal modification to GR, its capacity to incorporate admissible matter sources, and its potential to reconcile the existence of traversable wormholes with the known energy conditions - an outcome unattainable in standard GR.
\par A key mathematical feature of squared trace-extended gravity is its field equations, which include higher-order corrections to matter content. In this theory, the effective Einstein field equations are \cite{Roshan}: $\mathds{G}_{\mu \nu} = 8\pi \mathds{T}^{\mathfrak{eff}}_{\mu \nu},$
where the effective energy-momentum tensor $\mathds{T}^{\mathfrak{eff}}_{\mu \nu}$ includes both standard and trace-corrected contributions. The addition of quadratic $\mathds{T}$ terms introduces further nonlinearities into the field equations, necessitating a careful choice of the redshift and shape functions to ensure solvability. These nonlinearities also open the possibility of anisotropic pressures that are compatible with energy conditions, an outcome that is not easily achievable in GR.
\par One persuasive reason for exploring squared trace terms is the field of cosmology. It has been established that theories of this form can cause cosmic acceleration at later times and prevent Big Bang singularities \cite{Moraes2016}. This dual utility in both cosmological and astrophysical contexts attests to the theoretical strength of the squared trace gravity.
\par In short, traversable wormholes in squared trace-extended gravity are an exciting route for achieving stable and geometrically consistent wormhole geometries without the need to invoke exotic matter \cite{Tripathy, Zubair2021}. This theoretical development holds new promise for the physics of wormholes and further enlightens our comprehension of how gravity could behave outside the reach of GR.
\par New progress in Finsler geometry, especially within the framework of the Barthel connection formalism, presents an attractive context for investigating space-time anisotropies \cite{Hama1, Nekouee, Hama2022}. Finslerian generalizations of gravity theories are a natural arena for incorporating direction-dependent phenomena that do not exist in Riemannian geometry per se. Here, the squared trace alternative theory of gravity may be supplemented by Barthel-Finsler structures to encapsulate more general wormhole geometries that can incorporate anisotropic pressure distributions and non-trivial topologies. With this motivation in mind, the conventional $\mathfrak{f}(\mathds{R},\mathds{T})$ wormhole analysis can be extended to the Barthel-Finsler context, wherein the connection and curvature are no longer restricted under Riemannian symmetry. This geometric extension allows for the construction of wormhole solutions that are mathematically admissible and potentially provide a deeper understanding of matter-geometry couplings under anisotropic conditions; however, this area has not been fully explored in the current literature.
\par In summary, studying wormhole space-times does not replace black hole physics; instead, it enriches it. They provide a horizon-free, topologically nontrivial, anisotropic arena in which the role of energy conditions, matter-geometry coupling, and causal structure can be explored with unprecedented clarity. From this perspective, wormholes are not merely speculative geometries but indispensable tools for deducing the limits of classical general relativity and the viability of its extensions in the strong-gravity regime.
\par Despite these advances, wormhole solutions in Riemannian $\mathfrak{f}(\mathds{R},\mathds{T})$ and related modified gravity theories often remain viable only within narrow regions of the equation-of-state parameter space, particularly when classical energy conditions are imposed. In many cases, NEC-satisfying configurations require fine-tuned matter distributions or are restricted to specific choices of shape functions. This limitation motivates the exploration of geometric generalizations beyond the Riemannian framework. In this context, Finsler-Randers geometry offers a natural extension in which space-time anisotropy arises intrinsically from the geometry itself, providing a quantifiable enlargement of the admissible parameter space compared to its Riemannian baseline.
\par This paper investigates traversable wormholes in squared-trace extended gravity within a Finsler–Randers geometric framework equipped with the Barthel connection. Using the osculating Riemannian construction, we derive explicit analytical expressions for the energy density, radial and tangential pressures, anisotropy factor, and exoticity integral for two distinct wormhole shape functions. In contrast to previous studies that considered either Riemannian squared-trace gravity or Finslerian wormhole geometries separately, the present framework explicitly demonstrates how Finslerian anisotropy modifies the effective gravitational sector through the osculating metric. As a consequence, the admissible parameter space for non-exotic traversable wormholes is significantly enlarged, allowing physically viable solutions in parameter regions that are not admissible within the corresponding Riemannian \(\mathfrak{f}(\mathds{R},\mathds{T})=\mathds{R}+\ell_1\mathds{T}+\ell_2\mathds{T}^2\) framework.

\par The remainder of this paper is organized. In Sec.~\ref{sec2}, we introduce the geometric features of Barthel connections in Finsler spaces. Sec.~\ref{sec3}  describes the development of the basic framework of $\mathfrak{f}(\mathds{R},\mathds{T})$ gravity and the associated field equations. The wormhole configurations and energy condition analyses using the two shape functions in the Finsler framework are detailed in Secs.~\ref{sec4} and \ref{sec5}. Sec.~\ref{sec6} highlights the role of exotic matter in maintaining traversable wormholes, while Sec.~\ref{sec7} presents concluding remarks.
\section{Geometric Structures Induced by Barthel Connections in Finsler Spaces}\label{sec2}

Consider a smooth Finsler metric \(\mathbb{F}\) on a differentiable manifold \(\mathbb{M}\), equipped with local coordinates \((x^i, y^i)\) on the tangent bundle \(\mathbb{TM}\). The components of the metric tensor are expressed as:
\begin{equation}\label{Eq1}
\mathfrak{g}_{\iota\varsigma}=\mathbb{F}\mathbb{F}_{y^{\iota} y^{\varsigma}}+\mathbb{F}_{y^{\iota}}\mathbb{F}_{y^{\varsigma}},
\end{equation}
with \(\mathbb{F}_{y^{\iota}}=\frac{\partial \mathbb{F}}{\partial y^{\iota}}\).

In the realm of Finsler geometry, metrics of the \((\alpha, \beta)\)-type represent spaces constructed from a Riemannian metric \(\alpha(x,y)\) combined with a 1-form \(\beta(x,y)\). This setup differentiates them from traditional Riemannian geometries. Notable instances include Randers and Kropina metrics \cite{Matsumoto, Sabau}. The Riemannian element \(\alpha\) captures isotropic, spherically symmetric aspects, whereas the 1-form \(\beta\) incorporates directional variations \cite{Stavrinos}. The Randers metric takes the form:
\begin{equation}\label{Eq2}
\mathbb{F}(x,y)=\alpha(x,y)+\beta(x,y)=\sqrt{\mathfrak{a}_{\iota\varsigma} y^\iota y^\varsigma}+\mathbb{A}_\upsilon(x)y^\upsilon.
\end{equation}

This research aims to explore Finslerian applications in cosmological scenarios, emphasizing wormhole structures via Randers formalism. Notably, spatial coordinates are treated as comoving, and the time coordinate aligns with proper time observed by comoving entities. Inserting Eq.~(\ref{Eq2}) into Eq.~(\ref{Eq1}), assuming \(\mathbb{L}=\frac{1}{2}\mathbb{F}^2\), produces:
\begin{eqnarray}\label{Eq3}
\mathfrak{g}_{\iota\varsigma}(x,y)&=&\frac{\mathbb{L}_\alpha}{\alpha}\mathfrak{a}_{\iota\varsigma}+\frac{\mathbb{L}_{\alpha\alpha}}{\alpha^2}y_\iota y_\varsigma +\frac{\mathbb{L}_{\alpha\beta}}{\alpha}(\mathbb{A}_\iota y_\varsigma+\mathbb{A}_\varsigma y_\iota)+
\mathbb{L}_{\beta\beta}\mathbb{A}_\iota\mathbb{A}_\varsigma -\frac{\mathbb{L}_\alpha}{\alpha^3}y_\iota y_\varsigma.
\end{eqnarray}

Similar to Riemannian settings, covariant derivatives in Finsler geometry rely on connections. However, Finsler manifolds support multiple connection types, highlighting their structural diversity \cite{Matsumoto, Ingarden1, Ingarden2, Ingarden3}. This variety facilitates advanced geometric studies but poses challenges for cohesive theoretical models due to computational and interpretative complexities.

The Barthel connection was developed to address limitations in the Berwald connection and refine elements of the Cartan connection \cite{Ingarden1}. The configuration \((\mathbb{M}, \mathbb{F}(x,y), \mathbb{Y}(x))\) includes a Finsler manifold \((\mathbb{M}, \mathbb{F}(x,y))\) paired with a vector field \(\mathbb{Y}(x)\) on \(\mathbb{M}\). Requiring \(\mathbb{Y}(x) \neq 0\) everywhere in \(\mathbb{M}\) transforms the induced metric into a \(\mathbb{Y}\)-Riemannian form. The fundamental tensor yields the \(\mathbb{Y}\)-tensor field, defining the \(\mathbb{Y}\)-Riemannian structure. Compatible with the Finsler metric, the Barthel connection features a torsion tensor that maintains vector magnitudes during parallel translation, proving beneficial for physical applications. It enables viewing Finsler manifolds as point-wise structures with local Minkowski-like (non-Euclidean) properties. For \((\alpha,\beta)\)-based Finsler metrics, the Barthel linear connection-derived from the Cartan connection-functions as the Levi-Civita connection for the associated \(\mathbb{Y}\)-Riemannian manifold \cite{Hama}. This setup also connects to the linear \(\mathbb{Y}\)-connection, formed via the Cartan connection and a smooth, nonzero vector field \(\mathbb{Y}(x)\) (further details available in \cite{Hama, Yashwanth, Praveen3}).

Next, we examine the osculating Riemannian metric stemming from the Finsler metric \((\mathbb{M}, \mathbb{F})\). As previously noted, primary Finsler geometric elements reside on the total space \(\mathbb{TM}\) of the tangent bundle, with projection \(\pi_\mathbb{M}: \mathbb{TM} \rightarrow \mathbb{M}\). This space forms a \(2n\)-dimensional smooth manifold, described by local coordinates \((x^i, y^i)\). Here, \(x = (x^i)\) and \(y = (y^i)\) are independent variables. For example, the metric tensor \(\mathfrak{g}_{ij}: \mathbb{TM} \setminus \mathbb{O} \rightarrow \mathbb{R}\) is specified on the tangent bundle excluding the zero section \(\mathbb{O}\). The projection \(\pi_\mathbb{M}: \mathbb{TM} \to \mathbb{M}\) establishes a fiber bundle, allowing a local section \(\mathbb{Y}: \mathbb{U} \rightarrow \mathbb{TU}\), where \(\mathbb{U} \subset \mathbb{M}\) is open and \(\mathbb{Y}(x) \neq 0\) for \(x \in \mathbb{U}\). This satisfies \(\pi_\mathbb{M}(\mathbb{Y}(x)) = x\). The section \(\mathbb{Y}\) facilitates pulling back geometric objects from \(\mathbb{TM}\) to \(\mathbb{M}\). Thus, \(\mathfrak{g}_{ij} \circ \mathbb{Y}\) acts as a tensor on \(\mathbb{U}\), enabling the definition:
\begin{equation}\label{Eq4}
\mathfrak{g}_{ij}(x):=\mathfrak{g}_{ij}(x,y)|_{y=\mathbb{Y}(x)} ~~~ x\in\mathbb{U}.
\end{equation}
The pair \((\mathbb{U}, \mathfrak{g}_{ij})\) constitutes a Riemannian manifold, where \(\mathfrak{g}_{ij}(x) = \mathfrak{g}_{ij}(x, \mathbb{Y}(x))\) denotes the \(\mathbb{Y}\)-osculating Riemannian metric derived from the Finsler manifold \((\mathbb{M}, \mathbb{F})\).
\section{Conceptual Framework}\label{sec3}
Within a geometrically modified extended theory, we consider an action featuring a non-minimal coupling between matter and curvature, expressed as
\begin{equation}\label{Eq3.1}
\mathcal{S}=\int d^4x\sqrt{-\mathfrak{g}}\left(\frac{1}{2\varkappa^2}\mathfrak{f}(\mathds{R},\mathds{T})+\mathcal{L}_\mathfrak{m}\right).
\end{equation}
\par In this setup, $\mathcal{L}_\mathfrak{m}$ is the matter Lagrangian, and $\varkappa^2 = 8\pi$. Possible forms of the matter Lagrangian include $\mathcal{L}_\mathfrak{m}=-\mathds{T}$, $\mathcal{L}_\mathfrak{m}=-\rho$, and $\mathcal{L}_\mathfrak{m}=\mathfrak{p}$, with $\mathfrak{p}=\frac{p_r+2p_t}{3}$ representing the average pressure. Here, we consider $\mathcal{L}_\mathfrak{m}=\mathfrak{p}$ as it provides a natural and physically consistent choice. The connection between the matter Lagrangian and the energy-momentum tensor arises through the variational principle, leading to the expression,
\begin{equation}\label{Eq3.2}
\mathds{T}_{\mu\nu}=-\frac{2}{\sqrt{-\mathfrak{g}}}\frac{\delta(\sqrt{-\mathfrak{g}}\mathcal{L}_\mathfrak{m})}{\delta \mathfrak{g}^{\mu\nu}}.
\end{equation}
\par The action is generalized by introducing a function $\mathfrak{f}(\mathds{R}, \mathds{T})$ in place of the Ricci scalar $\mathds{R}$, with $\mathds{T}$ being the trace of the energy-momentum tensor. For simplicity and to preserve minimal coupling, the function $\mathfrak{f}(\mathds{R},\mathds{T})$ is commonly expressed as the sum $\mathfrak{f}(\mathds{R},\mathds{T})=\mathfrak{f}_1(\mathds{R})+\mathfrak{f}_2(\mathds{T})$, separating curvature and matter dependencies. By performing a variation of the modified gravitational action with respect to $\mathfrak{g}_{\mu\nu}$, one obtains the field equations for the extended theory as follows \cite{Parsaei,Tripathy}
\begin{eqnarray}\label{Eq3.3}
\mathds{R}_{\mu\nu}-\frac{\mathfrak{f}_1(\mathds{R})}{2\mathfrak{f}_{1,1}(\mathds{R})}\mathfrak{g}_{\mu\nu}&=&
\frac{1}{\mathfrak{f}_{1,1}(\mathds{R})}\bigg[\left(\nabla_\mu\nabla_\nu-\mathfrak{g}_{\mu\nu}\Box\right)\mathfrak{f}_{1,1}(\mathds{R})+\left(\varkappa^2
+\mathfrak{f}_{2,2}(\mathds{T})\right)
\mathds{T}_{\mu\nu}\nonumber\\&+&\left(\mathfrak{f}_{2,2}(\mathds{T})\mathfrak{p}+\frac{1}{2}\mathfrak{f}_{2}(\mathds{T})\right)\mathfrak{g}_{\mu\nu}\bigg],
\end{eqnarray}
for convenience, we introduced abbreviated notation for the partial derivatives with respect to the Ricci scalar $\mathds{R}$ and the trace $\mathds{T}$ of the energy-momentum tensor as follows,
\begin{equation}\label{Eq3.4}
\mathfrak{f}_{1,1}(\mathds{R})=\frac{\partial \mathfrak{f}_{1}(\mathds{R})}{\partial \mathds{R}}, ~~~ \mathfrak{f}_{2,2}(\mathds{T})=\frac{\partial \mathfrak{f}_{2}(\mathds{T})}{\partial \mathds{T}}.
\end{equation}
We assume $\mathfrak{f}_1(\mathds{R})=\mathds{R}$, such that the corresponding modified theory mimics GR in its field equations
\begin{equation}\label{Eq3.5}
\mathds{G}_{\mu\nu}=(\varkappa^2+\mathfrak{f}_{2,2}(\mathds{T}))
\mathds{T}_{\mu\nu}+\left(\mathfrak{f}_{2,2}(\mathds{T})\mathfrak{p}+\frac{1}{2}\mathfrak{f}_{2}(\mathds{T})\right)\mathfrak{g}_{\mu\nu},
\end{equation}
by rewriting Eq.~(\ref{Eq3.5}), we obtain
\begin{equation}\label{Eq3.6}
\mathds{G}_{\mu\nu}=\varkappa^2_{\mathds{T}}\left(\mathds{T}_{\mu\nu}+\mathds{T}^{\mathfrak{int}}_{\mu\nu}\right),
\end{equation}
where $\varkappa^2_{\mathds{T}}=\varkappa^2+\mathfrak{f}_{2,2}(\mathds{T})$, $\mathds{G}_{\mu\nu}$ represents the Einstein tensor, with an interaction term that is expressed as follows,
\begin{equation}\label{Eq3.7}
\mathds{T}^{\mathfrak{int}}_{\mu\nu}=\left(\frac{\mathfrak{f}_{2,2}(\mathds{T})\mathfrak{p}+\frac{1}{2}\mathfrak{f}_{2}(\mathds{T})}{\varkappa^2
+\mathfrak{f}_{2,2}(\mathds{T})}\right)\mathfrak{g}_{\mu\nu}.
\end{equation}
Owing to modifications in the geometry of the action, the interaction term may originate from quantum effects or imperfect fluids. Its presence implies that the non-minimal coupling between matter and geometry induces an effective matter field, which might be responsible for the late-time acceleration of the universe. It is worth noting that adopting a linear form for the functional $\mathfrak{f}_2(\mathds{T})$ yields a constant, redefined Einstein constant $\varkappa_{\mathds{T}}^{2}$.
In contrast, the present study explores a quadratic form, $\mathfrak{f}_2(\mathds{T})=\ell_1\mathds{T}+\ell_2\mathds{T}^2$, which results in a $\mathds{T}$-dependent Einstein constant. Here, $\ell_1$ and $\ell_2$ are model parameters that will be constrained based on the specific wormhole solutions under consideration. Under this choice, the interaction term takes the form,
\begin{equation}\label{Eq3.8}
\mathds{T}^{\mathfrak{int}}_{\mu\nu}=\frac{1}{\varkappa_{\mathds{T}}^{2}}\left[\ell_1\left(\mathfrak{p}+\frac{\mathds{T}}{2}\right)
+\ell_2\mathds{T}\left(2\mathfrak{p}+\frac{\mathds{T}}{2}\right)\right]\mathfrak{g}_{\mu\nu}.
\end{equation}
\par Traversable wormhole solutions are typically formulated under the assumption of an anisotropic fluid, described by the energy-momentum tensor,
\begin{equation}\label{Eq3.9}
\mathds{T}_{\mu}^{\nu}=\textrm{diag}\left(\rho,-p_r,-p_t,-p_t\right),
\end{equation}
where $p_r$ and $p_t$ are the radial and tangential pressure components, respectively, and $\rho$ is the energy density associated with the matter content, and the trace of the energy-momentum tensor is $\mathds{T}=\rho-p_r-2p_t$.
\section{Finslerian Perspectives on Wormhole Geometry}\label{sec4}
To explore wormhole configurations within the framework of Finsler geometry using the Barthel connection, we integrate the vector field tied to the Randers structure into the general expression of the Finslerian metric. Beginning with a wormhole solution derived from GR, we isolate its Riemannian component and introduce a 1-form to construct the corresponding Finsler-Randers geometry. Applying the osculating method to this framework yields a Barthel-type Finsler-Randers wormhole model that exhibits a Lorentzian signature. The Riemannian metric component $\alpha(x,y)$ represents spatial isotropy, while the 1-form $\beta(x,y)$ introduces directional dependence, indicating anisotropy or potential Lorentz symmetry breaking \cite{Matsumoto, Sabau}. Consequently, the Riemannian metric tensor $\mathfrak{a}_{ij}$ takes the form,
\begin{equation}\label{Eq14}
\mathfrak{a}_{ij}(x)=\textrm{diag}\left(\textrm{e}^{2\Psi(r)},-\frac{1}{1-\frac{b(r)}{r}},-r^2,-r^2\sin^2\theta\right),
\end{equation}
based on the assumptions established in the section II, we now focus on an \((\alpha, \beta)\)-type metric and analyze the scenario where the vector field \( \mathds{Y} = \mathds{A} \) is specified, with its components expressed as \( \mathds{A}^i = \mathfrak{a}^{ij} \mathds{A}_j \). The requirement that the vector field $\mathds{A}$ remains non-zero throughout the manifold $\mathds{M}$ implies that the 1-form $\beta$ has no vanishing points.
With this framework, we introduce the concept of the $\mathds{A}$-osculating Riemannian manifold, denoted by \( (\mathds{M}, \mathfrak{g}_{ij}) \), where the metric tensor is defined as \( \mathfrak{g}_{ij}(x) := \mathfrak{g}_{ij}(x, \mathds{A}) \).
Let $\Im$ represent the norm of the vector field $\mathds{A}$ measured with respect to the Riemannian metric 
$\alpha$, which yields the following relation,
\begin{eqnarray}\label{Eq15}
&~&\mathds{Y}_i(x, \mathds{A}) = \mathds{A}_i, ~~~ \Im^2 = \mathds{A}^i \mathds{A}_i = \alpha^2(x, \mathds{A}), \nonumber\\\nonumber\\&~& ~\beta(x,\mathds{A})=\mathds{A}_i\mathds{A}^i=\Im^2,
\end{eqnarray}
referring to Eq.~(\ref{Eq15}), one finds
\begin{equation}\label{Eq16}
\mathfrak{g}_{ij}(x,y)|_{y=\mathds{A}(x)}=(1+\Im)\mathfrak{a}_{ij}(x)+\left(\frac{1}{\Im}+1\right)\mathds{A}_{i}(x)\mathds{A}_{j}(x).
\end{equation}
Under the assumption that $(\mathds{A}_i)=(\xi(r),0,0,0)$, the Finsler-Barthel-Randers metric exhibits the following form
\begin{eqnarray*}
\alpha(x,y)|_{y=\mathds{A}(x)}&=&\xi(r),\\\\
\beta(x,y)|_{y=\mathds{A}(x)}&=&\xi^2(r),
\end{eqnarray*}
\begin{eqnarray}\label{Eq17}
\mathfrak{g}_{ij}(x,y)|_{y=\mathds{A}(x)}=\bigg(1+\xi(r)\bigg)\begin{bmatrix}
   \textrm{e}^{2\Psi(r)}+\xi(r) & 0 & 0 & 0 \\
    0 & -\frac{1}{1-\frac{b(r)}{r}} & 0 & 0 \\
    0 & 0 & -r^2 & 0\\
    0 & 0 & 0 & -r^2\sin^2\theta
\end{bmatrix},
\end{eqnarray}
with $i,j\in\{0,1,2,3\}$, corresponding to $\{t,r,\theta,\phi\}$, the functions \( \Psi(r) \) and $b(r)$ are interpreted as the redshift and shape functions, respectively \cite{Konoplya}. The redshift function \( \Psi(r) \) describes how redshift and tidal gravitational effects evolve with radial distance in the wormhole structure. To prevent the formation of horizons within the wormhole geometry, it is crucial that the redshift function 
$\Psi(r)$ remains finite throughout the space-time. The regularity of $\Psi(r)$ ensures the traversability of the wormhole, ruling out the presence of singularities or event horizons. A horizon, particularly in the case where $\Psi(r)=0$, would obstruct bidirectional passage through the wormhole \cite{Morris,Hohmann}. This case, referred to as a tideless condition, is significant because it implies the absence of substantial tidal forces within the wormhole \cite{Konoplya, Krishna, Sanjay}. Following the motivation in Ref. \cite{Yashwanth}, we consider $\xi(r)$ to be a constant. This assumption is primarily adopted for analytical tractability, as allowing $\xi(r)$ to vary would introduce additional derivative terms and nonlinear couplings in the field equations, making it difficult to obtain closed-form solutions in the present framework. The constant $\xi$ in the osculating Finsler-Barthel-Randers wormhole tensor modifies the metric components, influencing curvature, gravitational strength, and wormhole features such as stability, traversability, and particle motion. It also introduces anisotropy and may reduce the need for exotic matter. Variations in $\xi$ could lead to observable effects such as lensing or redshift, while in the limit \( \xi \to 0 \), the metric (\ref{Eq17}) reduces to the corresponding Riemannian wormhole case. A more general analysis with variable \( \xi \) is left for future work.

To derive the effective field equations in the Finsler–Barthel framework, we employ the osculating Riemannian procedure, in which the Finsler metric tensor is evaluated along a non-vanishing supporting vector field \(\mathds{A}(x)\). In this approach, the original Finslerian geometry is locally approximated by an effective Riemannian geometry whose metric coefficients depend on both the space-time coordinates and the chosen vector field. The resulting osculating metric is given by Eq.~(\ref{Eq17}).\\
Assuming a constant redshift function \(\Psi(r)=0\), the effective Finsler–Barthel wormhole line element takes the form
\begin{equation*}
	ds^2=(1+\xi)\left[(1+\xi)dt^2-\frac{dr^2}{1-b(r)/r}-r^2\left(d\theta^2+\sin^2\theta d\phi^2\right)\right].
\end{equation*}
The parameter \(\xi\) characterizes the Finslerian deformation and rescales the metric components relative to the standard Riemannian wormhole geometry. Using the Levi-Civita connection associated with the osculating metric, one can compute the corresponding Christoffel symbols, Ricci tensor \(\mathds{R}_{\mu\nu}\), Ricci scalar \(\mathds{R}\), and finally the Einstein tensor \(\mathds{G}_{\mu\nu}=\mathds{R}_{\mu\nu}-\frac{1}{2}\mathds{R}\mathfrak{g}_{\mu\nu}\). The non-vanishing components are obtained as
\begin{eqnarray*}
	\mathds{G}^{t}_{\ t}&=&\frac{b'(r)}{(1+\xi)r^2},\\
	\mathds{G}^{r}_{\ r}&=&\frac{b(r)}{(1+\xi)r^3},\\
	\mathds{G}^{\theta}_{\ \theta}&=&\mathds{G}^{\phi}_{\ \phi}=\frac{rb'(r)-b(r)}{(1+\xi)r^3}.
\end{eqnarray*}

\par The existence of traversable wormholes can be supported by a fluid whose radial and tangential pressures follow the barotropic relations 
$p_r=\varpi_1\rho$ and $p_t=\varpi_2\rho$, with $\varpi_1$ and $\varpi_2$ as constants of the EoS. These constants may be freely chosen or physically motivated, based on the desired scenario. By using Eqs. (\ref{Eq3.6}) and (\ref{Eq3.8}), the field equations for the traversable wormhole space-time become,
\begin{equation}\label{Eq17.1}
	\mathds{G}_{\mu\nu}=\left(8\pi+\ell_1+2\ell_2\mathds{T}\right)\mathds{T}_{\mu\nu}+	\left[\ell_1\left(\mathfrak{p}+\frac{\mathds{T}}{2}\right)
+\ell_2\mathds{T}\left(2\mathfrak{p}+\frac{\mathds{T}}{2}\right)\right]\mathfrak{g}_{\mu\nu}.
\end{equation}

The above construction explicitly shows how the Finslerian deformation parameter enters the curvature sector via the osculating metric and modifies the effective gravitational dynamics compared with the standard Riemannian case \cite{Tripathy}. Given such barotropic forms and Eq. (\ref{Eq17.1}), we have
\begin{eqnarray}
\frac{b^\prime(r)}{(\xi+1)r^2}&=&\overset{\lambda_1}{\left[\overbrace{8\pi+\frac{\ell_1}{6}(9-\varpi_1-2\varpi_2)}\right]}\rho\nonumber\\&+&\left[\frac{\ell_2}{6}(1-\varpi_1-2\varpi_2)\overset{\lambda_2}{(\overbrace{15+\varpi_1+2\varpi_2})}\right]\rho^2,\label{Eq18}\\
\nonumber\\\frac{b(r)}{(\xi+1)r^3}&=&\overset{\lambda_3}{\left[\overbrace{-8\pi\varpi_1+\frac{\ell_1}{6}(3-7\varpi_1-2\varpi_2)}\right]}\rho\nonumber\\&+&\left[\frac{\ell_2}{6}(1-\varpi_1-2\varpi_2)\overset{\lambda_4}{(\overbrace{3-11\varpi_1+2\varpi_2})}\right]\rho^2,\label{Eq19}\\
\nonumber\\\frac{rb^\prime(r)-b(r)}{(\xi+1)r^3}&=&\overset{\lambda_5}{\left[\overbrace{-16\pi\varpi_2+\frac{\ell_1}{3}(3-\varpi_1-8\varpi_2})\right]}\rho\nonumber\\&+&\left[\frac{\ell_2}{6}(1-\varpi_1-2\varpi_2)\overset{\lambda_6}{(\overbrace{6+2\varpi_1-20\varpi_2})}\right]\rho^2.\label{Eq20}
\end{eqnarray}
Using Eqs.~(\ref{Eq18}) and (\ref{Eq19}), one can obtain the expression for the energy density of the traversable wormhole,
\begin{equation}\label{Eq21}
\rho=\left[\frac{\lambda_4r\frac{b^\prime(r)}{b(r)}-\lambda_2}{\lambda_1\lambda_4-\lambda_2\lambda_3}\right]\frac{b(r)}{(\xi+1)r^3}.
\end{equation}
The energy density depends on both the EoS parameters and the chosen shape function of the wormhole geometry. The alternative theory of gravity under consideration introduces two arbitrary constants, $\ell_1$ and $\ell_2$. Constraints on $\ell_1$, $\varpi_1$, and $\varpi_2$ are derived directly from Eqs.~(\ref{Eq18}) through (\ref{Eq20}) as follows,
\begin{equation}\label{Eq22}
\lambda_1-\lambda_3=\lambda_5,
\end{equation}
\begin{equation}
\buildrel{\chi}\over{\overbrace{\frac{\ell_2}{6}(1-\varpi_1-2\varpi_2)}}\left(\lambda_2-\lambda_4\right)=\frac{\ell_2}{6}(1-\varpi_1-2\varpi_2)\lambda_6.\label{Eq23}
\end{equation}
By enforcing the condition (\ref{Eq22}), we establish a correspondence between $\varpi_1$ and $\varpi_2$ as following form,
\begin{equation}\label{Eq24}
\varpi_1=-(2\varpi_2+\gamma), ~~~ \gamma=\frac{6\pi}{\ell_1+6\pi}.
\end{equation}
With the trace defined as $\mathds{T}=\rho-p_r-2p_t=(1-\varpi_1-2\varpi_2)\rho$, a corresponding relation involving $\chi$ can be established as follows,
\begin{equation}\label{Eq25}
\chi=\frac{\ell_2}{6}\frac{\mathds{T}}{\rho}.
\end{equation}
The field equation constraints, particularly Eq.~(\ref{Eq23}) coupled with the requirement $\chi\neq0$, strictly impose $\ell_1=4\pi$. Alternatively, if the trace of the matter field were to vanish ($\chi=0$), the consistency condition would yield $\ell_1=-12\pi$. To avoid a vanishing trace in our configuration, we consistently choose $\ell_1=4\pi$, establishing that the model imposes rigid constraints on its parameter space rather than allowing arbitrary selections. It is interesting to note that the final reduced expression for the energy density of the traversable wormhole model does not explicitly depend on the quadratic trace coupling parameter \(\ell_2\), while the remaining parameter \(\ell_1\) becomes constrained through the consistency conditions of the field equations. Owing to the algebraic structure of the modified equations under the imposed barotropic equations of state, the explicit dependence on \(\ell_2\) also disappears from the corresponding pressure components and energy conditions. However, this does not imply that the underlying \(\mathfrak{f}(\mathds{R},\mathds{T})\) theory can be consistently reduced to a purely linear function of \(\mathds{T}\).\\
In the present framework, the squared-trace sector plays an essential structural role in maintaining nontrivial matter configurations with non-vanishing trace. In particular, removing the quadratic contribution by setting \(\ell_2=0\) leads to consistency conditions that force the trace parameter \(\chi\) to vanish, thereby reducing the model to an effectively traceless matter sector. To avoid such a traceless configuration and preserve the non-exotic matter regions obtained in this work, it is therefore necessary to retain a finite non-zero value of \(\ell_2\).

\section{Wormhole Configurations and their Compatibility with Energy Conditions}\label{sec5}
To study non-exotic traversable wormholes under the squared trace extended gravity framework, we select two forms of the shape function: an exponential type $b(r)=r\textrm{e}^{-2(r-r_0)}$, and a power-law type $b(r)=\sqrt{r_0r}$, where $r_0$ represents the radius at the wormhole throat. These choices are physically motivated and widely used in the wormhole literature, allowing direct comparison with earlier studies (e.g., Ref. \cite{Tripathy}). The exponential form describes a rapidly decaying geometry, leading to localized matter distributions and favorable asymptotic behavior, while the power-law form represents a more gradual variation, enabling the exploration of extended configurations. It is essential to emphasize that the chosen shape functions must meet specific geometric and physical criteria. These include reproducing the throat condition $b(r_0)=r_0$, ensuring that the derivative at the throat satisfies $b^\prime(r_0)\leq1$, and fulfilling both the metric requirement $1-\frac{b(r)}{r}\geq0$ $(r>r_0)$ and the flare-out condition $\frac{b(r)-rb^\prime(r)}{b^2(r)}>0$. 
\subsection{Case 1:}
\begin{figure*}[htpb]
\centering
\mbox{{\includegraphics[scale=0.26]{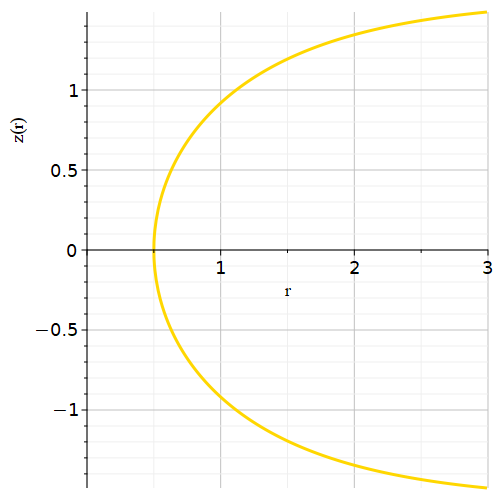}}
{\includegraphics[scale=0.35]{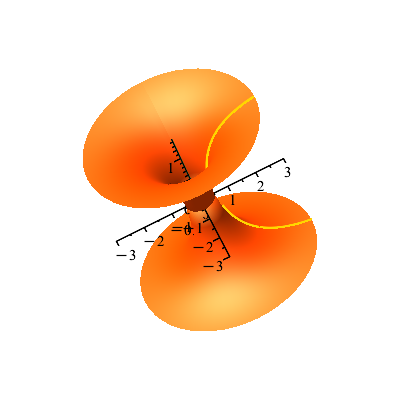}}{\includegraphics[scale=0.26]{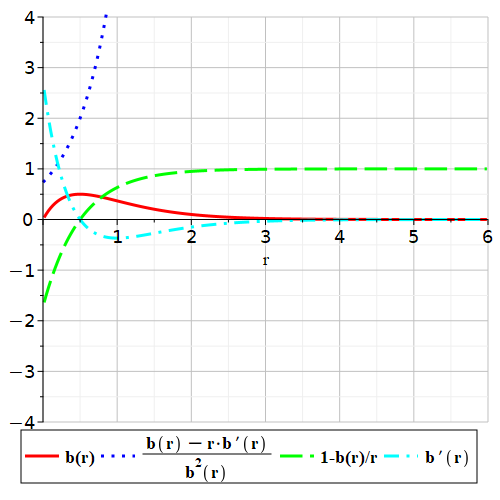}}}
\caption{\label{fig1} Embedding surfaces in two and three dimensions constructed from the exponential shape function $b(r)=r\textrm{e}^{-2(r-r_0)}$ and its behavior at $r_0 = 0.5$.}
\end{figure*}

We begin with the exponential shape function $b(r)=r\textrm{e}^{-2(r-r_0)}$. The derivative is found to be $b^\prime(r)=(1-2r)\frac{b(r)}{r}$, yielding 
$\frac{b^\prime(r)}{b(r)}=\frac{1}{r}-2$. Figure~\ref{fig1} shows that the function and its derivative meet all the geometric criteria required for a physically consistent wormhole and illustrates the embedding surfaces of the wormhole in two and three dimensions, constructed using this shape function.
By employing this shape function, one obtains the energy density, radial pressure, and tangential pressure as,
\begin{eqnarray}
\rho&=&\left[\frac{-2r(3-11\varpi_1+2\varpi_2)-12(\varpi_1+1)}{(2\varpi_1\varpi_2+2\varpi_2+\varpi_1^2+3)(8\pi+3\ell_1)-12\ell_1+32\pi\varpi_1}\right]\nonumber\\&\times&\frac{1}{(\xi+1)r^2\textrm{e}^{2(r-r_0)}},\label{Eq26}\\
\nonumber\\p_r&=&\left[\frac{2r(3-11\varpi_1+2\varpi_2)+12(\varpi_1+1)}{(2\varpi_1\varpi_2+2\varpi_2+\varpi_1^2+3)(8\pi+3\ell_1)-12\ell_1+32\pi\varpi_1}\right]\nonumber\\&\times&\frac{2\varpi_2+\gamma}{(\xi+1)r^2\textrm{e}^{2(r-r_0)}},\label{Eq27}\\
\nonumber\\p_t&=&\left[\frac{-2r(3-11\varpi_1+2\varpi_2)-12(\varpi_1+1)}{(2\varpi_1\varpi_2+2\varpi_2+\varpi_1^2+3)(8\pi+3\ell_1)-12\ell_1+32\pi\varpi_1}\right]\nonumber\\&\times&\frac{\varpi_2}{(\xi+1)r^2\textrm{e}^{2(r-r_0)}}.\label{Eq28}
\end{eqnarray}
This work aims to investigate the existence of traversable wormholes that do not require exotic matter within the context of extended gravity theories in the Finsler framework. The present study has examined several scenarios, including one where $\varpi_2$ is set to $-0.465$ (Negative, $>-1$	Quintessence-like in tangential direction), resulting in $\varpi_1$ taking the value $0.33$ (Positive, $\approx 1/3$ Radiation-like in radial direction). Based on the assumptions stated above and taking $\xi<-1$, the energy density is found to be positive. We now proceed to examine the energy conditions for a wormhole characterized by an exponential shape function as follows,
\begin{itemize}
\item NEC: 
\begin{eqnarray}
\rho+p_r&=&\left[\frac{-2r(3-11\varpi_1+2\varpi_2)-12(\varpi_1+1)}{(2\varpi_1\varpi_2+2\varpi_2+\varpi_1^2+3)(8\pi+3\ell_1)-12\ell_1+32\pi\varpi_1}\right]\nonumber\\&\times&\frac{1-(2\varpi_2+\gamma)}{(\xi+1)r^2\textrm{e}^{2(r-r_0)}},\label{Eq29}\\
\rho+p_t&=&\left[\frac{-2r(3-11\varpi_1+2\varpi_2)-12(\varpi_1+1)}{(2\varpi_1\varpi_2+2\varpi_2+\varpi_1^2+3)(8\pi+3\ell_1)-12\ell_1+32\pi\varpi_1}\right]\nonumber\\&\times&\frac{1+\varpi_2}{(\xi+1)r^2\textrm{e}^{2(r-r_0)}}.\label{Eq30}
\end{eqnarray}
\begin{equation*}
\begin{cases}
\rho+p_r=1.33\rho\geq0,\\
\rho+p_t=0.535\rho\geq0,
\end{cases}
 \Rightarrow ~~ \textrm{NEC is satisfied}
\end{equation*}
\item Weak Energy Condition (WEC):
\begin{equation*}
\begin{cases}
\rho\geq0,\\
 \rho+p_i\geq0 ~~\forall i=\{r,t\}.
\end{cases}
 \Rightarrow ~~ \textrm{WEC is satisfied}
\end{equation*}
\item Strong Energy Condition (SEC):
\begin{eqnarray}\label{Eq31}
\rho+p_r+2p_t&=&\left[\frac{-2r(3-11\varpi_1+2\varpi_2)-12(\varpi_1+1)}{(2\varpi_1\varpi_2+2\varpi_2+\varpi_1^2+3)(8\pi+3\ell_1)-12\ell_1+32\pi\varpi_1}\right]\nonumber\\&\times&\frac{1-\gamma}{(\xi+1)r^2\textrm{e}^{2(r-r_0)}},
\end{eqnarray}
\begin{equation*}
\begin{cases}
 \rho+p_i\geq0 ~~\forall i=\{r,t\},\\
\rho+p_r+2p_t=(1+0.08-0.68)\rho=0.4\rho\geq0.
\end{cases}
 \Rightarrow ~~ \textrm{SEC is satisfied}
\end{equation*}
\item Dominant Energy Condition (DEC):
\begin{eqnarray}
\rho-p_r&=&\left[\frac{-2r(3-11\varpi_1+2\varpi_2)-12(\varpi_1+1)}{(2\varpi_1\varpi_2+2\varpi_2+\varpi_1^2+3)(8\pi+3\ell_1)-12\ell_1+32\pi\varpi_1}\right]\nonumber\\&\times&\frac{1+(2\varpi_2+\gamma)}{(\xi+1)r^2\textrm{e}^{2(r-r_0)}},\label{Eq32}\\
\rho-p_t&=&\left[\frac{-2r(3-11\varpi_1+2\varpi_2)-12(\varpi_1+1)}{(2\varpi_1\varpi_2+2\varpi_2+\varpi_1^2+3)(8\pi+3\ell_1)-12\ell_1+32\pi\varpi_1}\right]\nonumber\\&\times&\frac{1-\varpi_2}{(\xi+1)r^2\textrm{e}^{2(r-r_0)}}.\label{Eq33}
\end{eqnarray}
\begin{equation*}
\begin{cases}
\rho\geq|p_r|=0.33\rho,\\
\rho\geq|p_t|=0.465\rho.
\end{cases}
\Rightarrow ~~ \textrm{DEC is satisfied}
\end{equation*}
\end{itemize}
\begin{figure*}[htpb]
\centering
\mbox{{\includegraphics[scale=0.24]{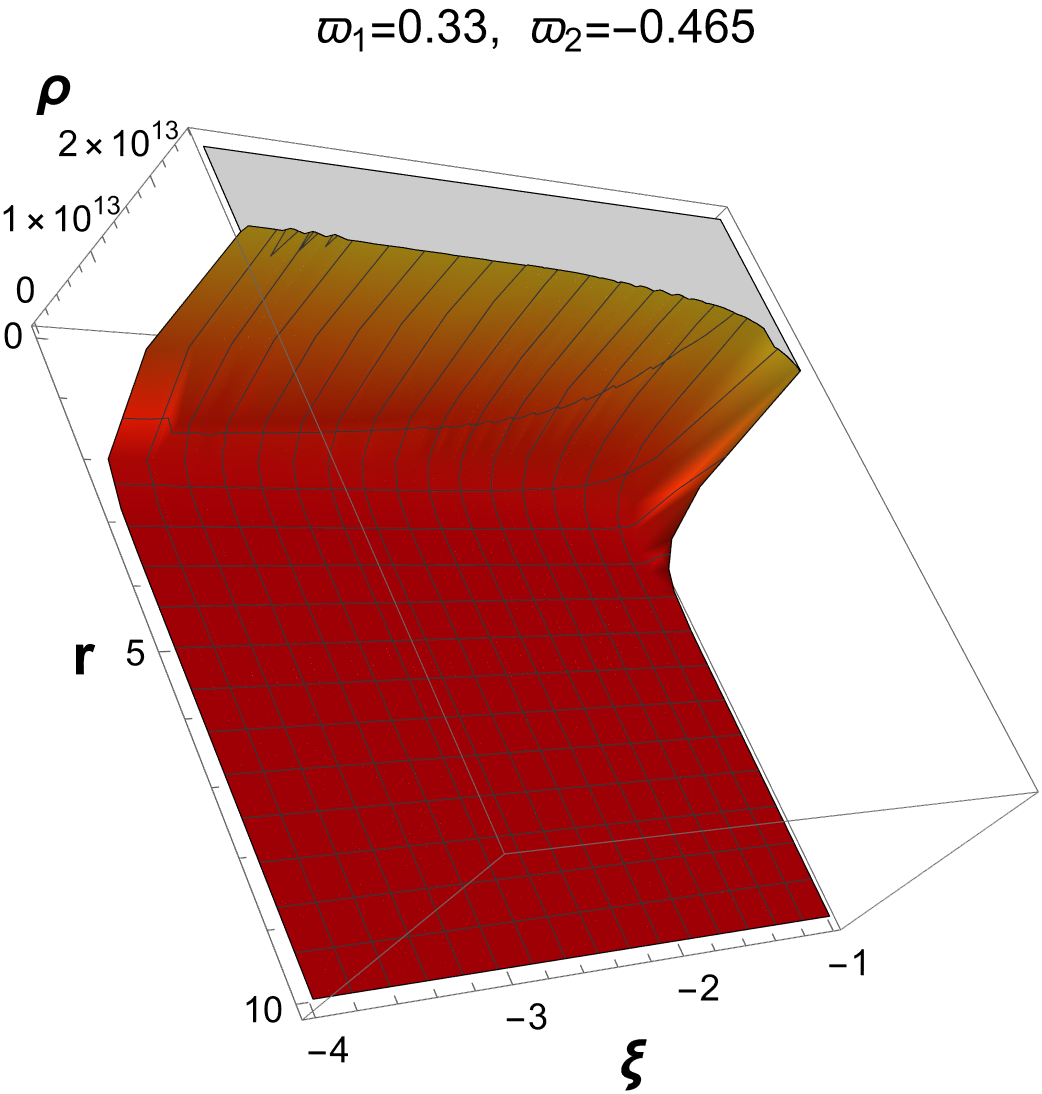}}
{\includegraphics[scale=0.26]{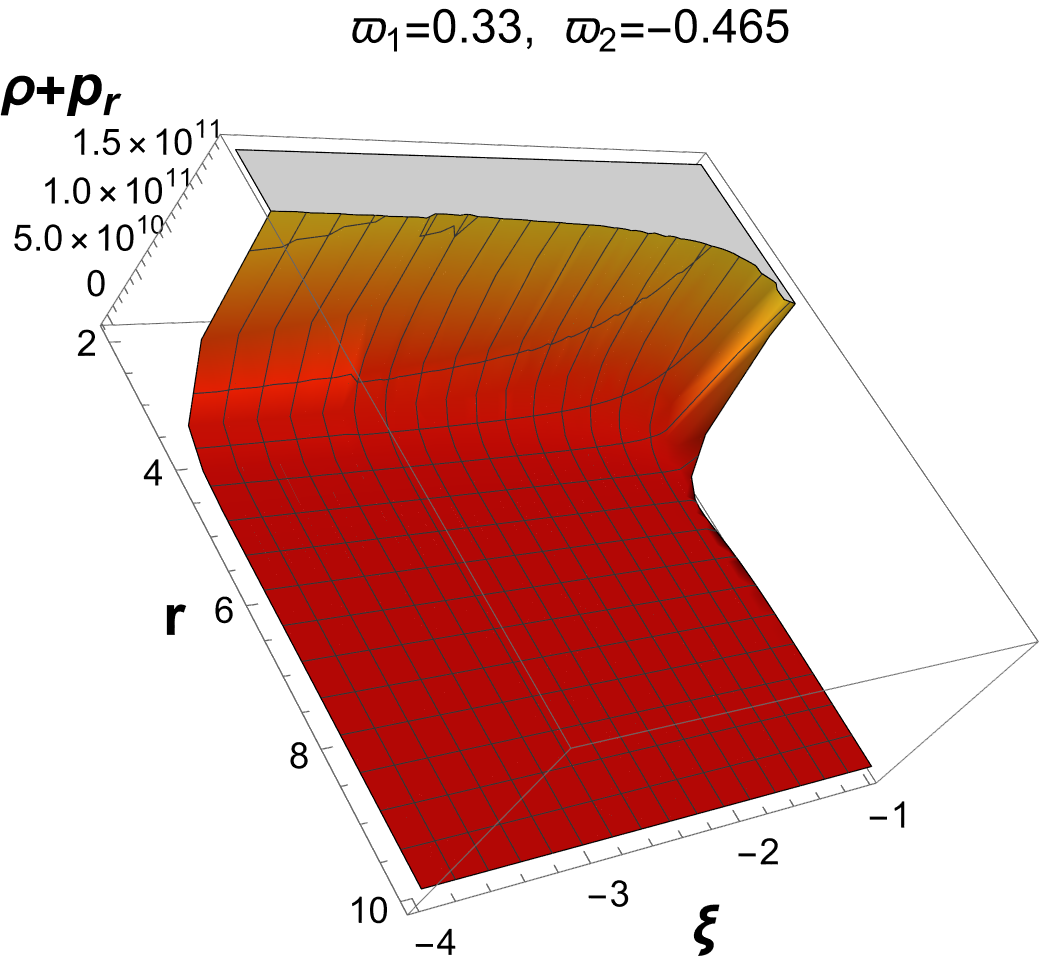}}{\includegraphics[scale=0.26]{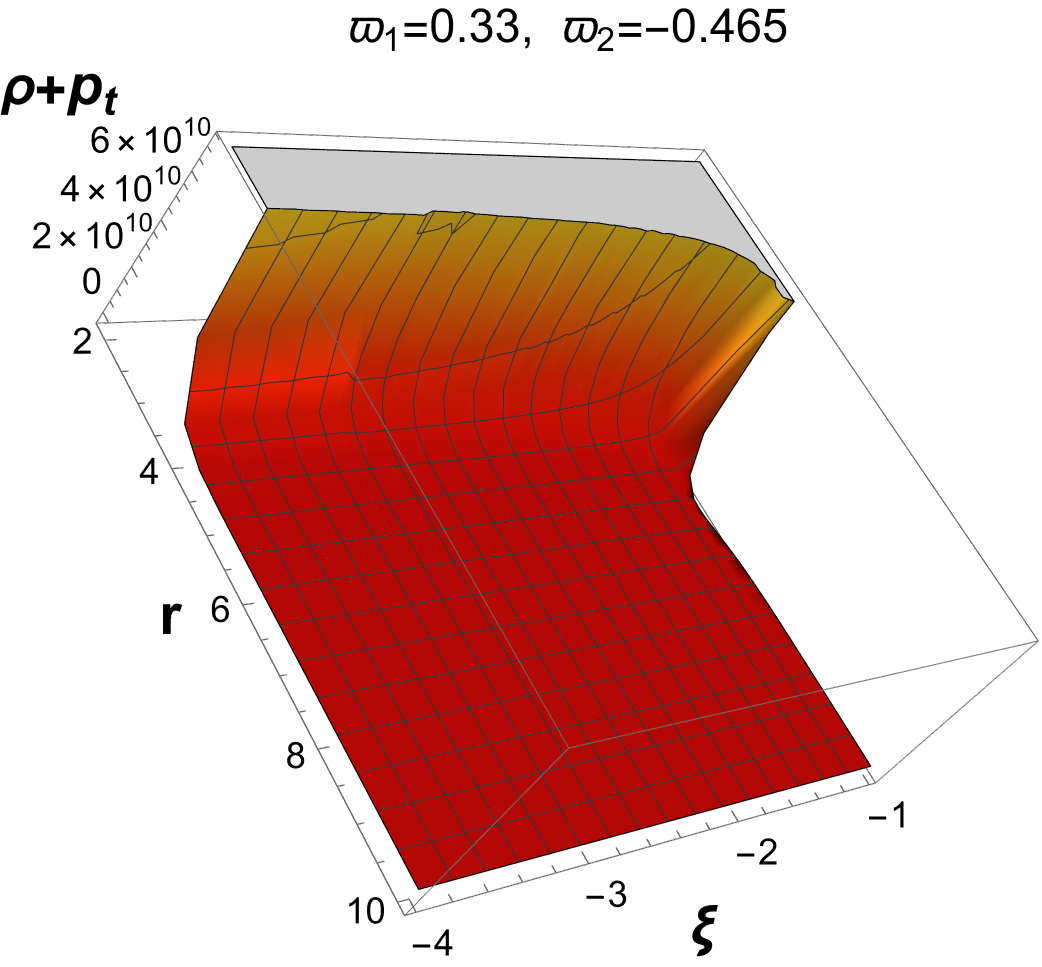}}}\\
\mbox{{\includegraphics[scale=0.29]{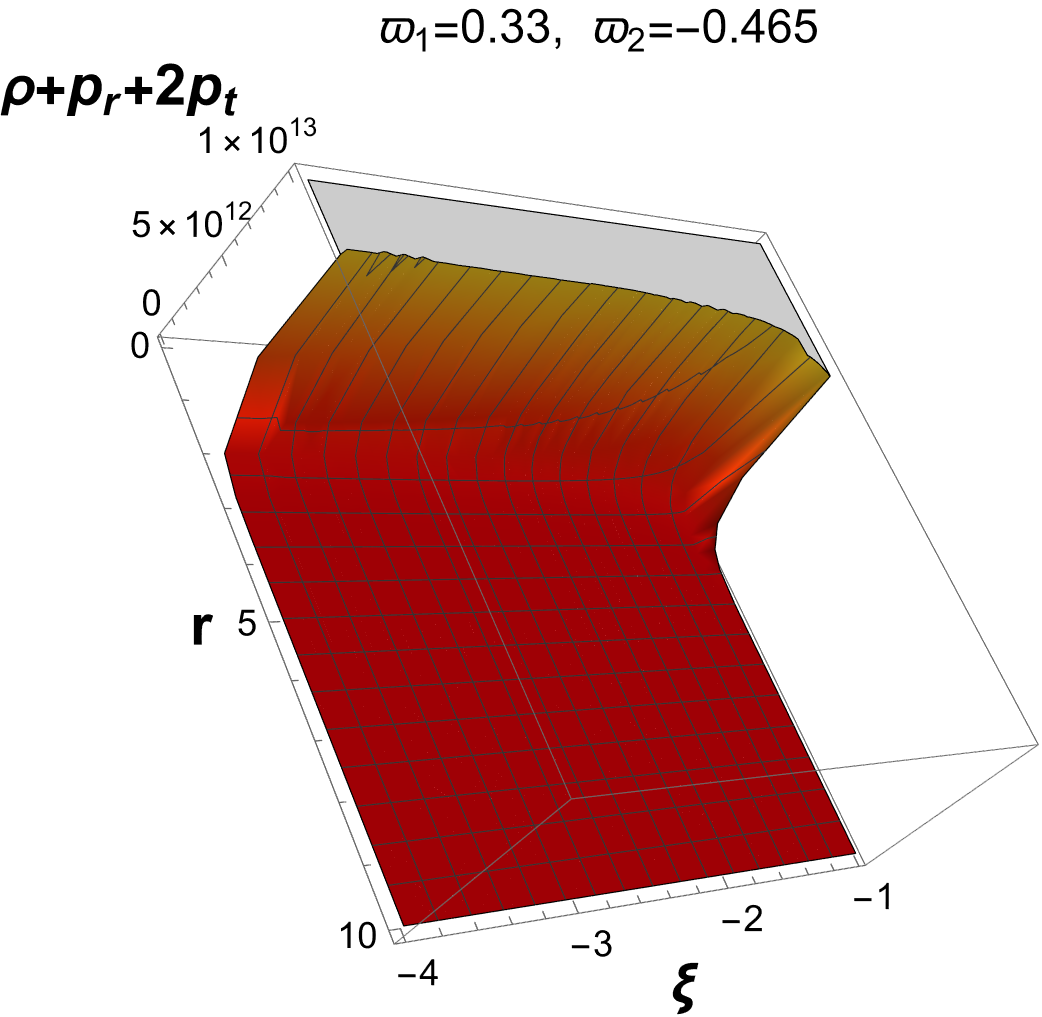}}
{\includegraphics[scale=0.26]{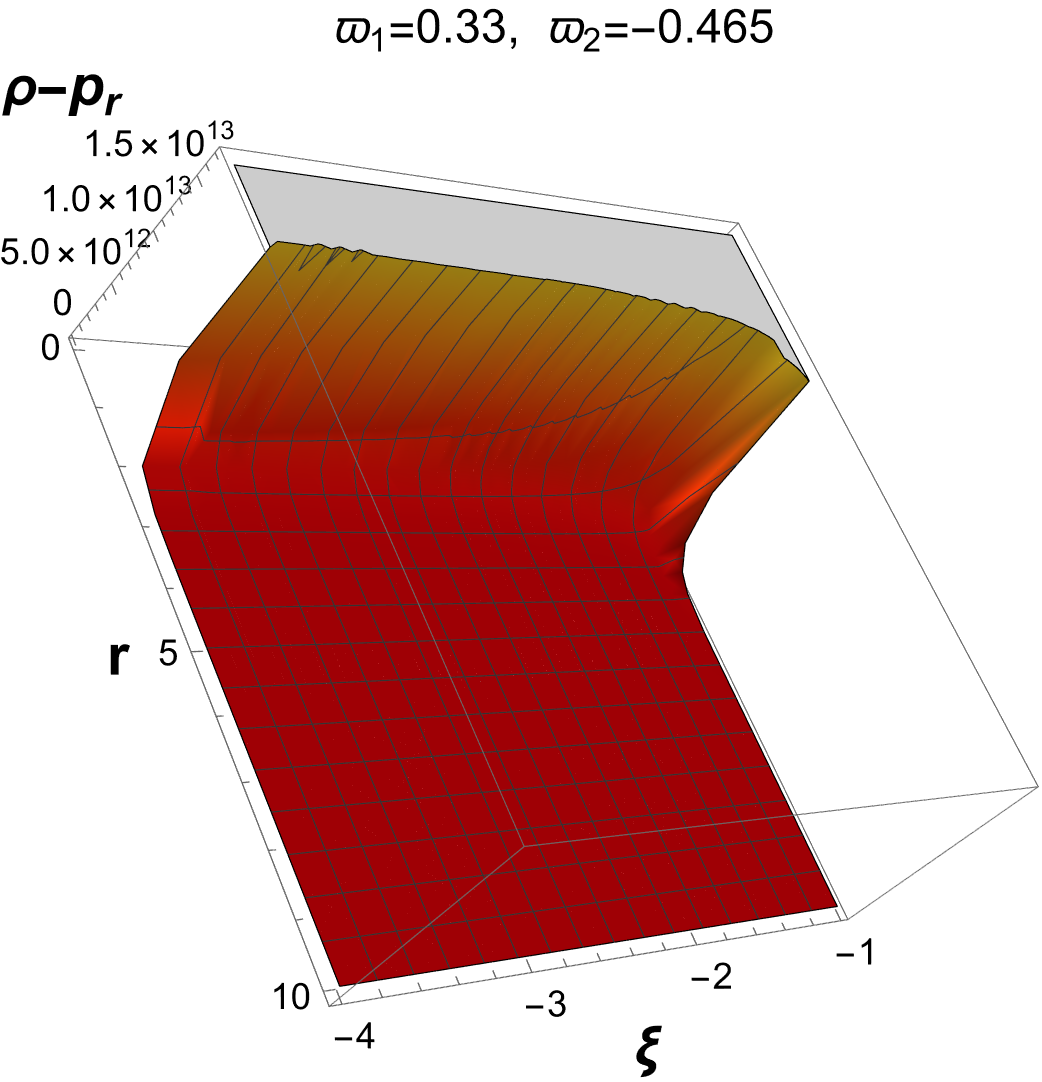}}{\includegraphics[scale=0.26]{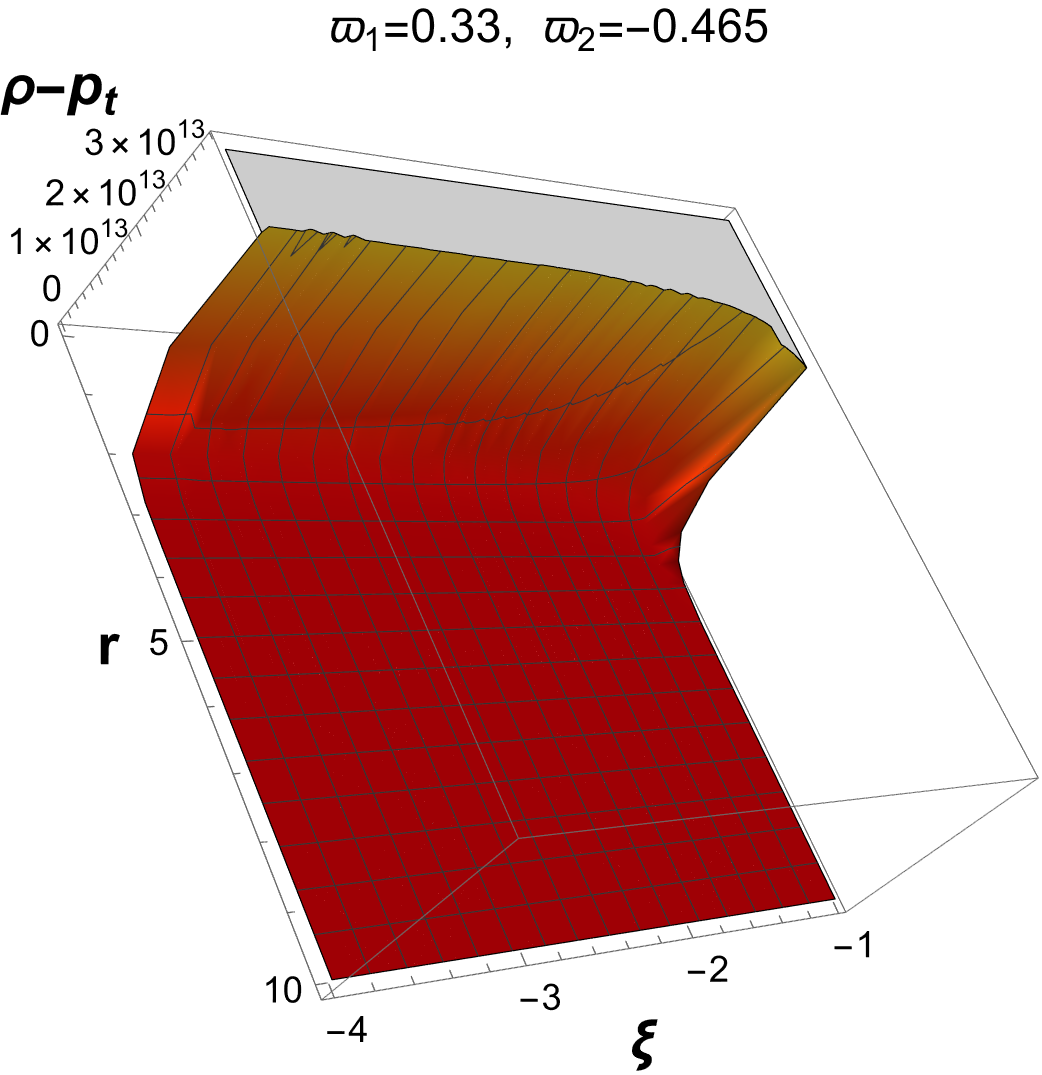}}}
\caption{\label{fig2} Satisfaction of all energy conditions (NEC, WEC, SED, DEC) at $r_0=0.5$ and $\xi<-1$.}
\end{figure*}
The results obtained in Fig.~\ref{fig2} are significant because they enable a wormhole configuration that does not violate the NEC, indicating that such a structure could be supported by anisotropic, non-exotic matter exhibiting a combination of quintessence-like and radiation-like behavior. May be interpreted phenomenologically as an anisotropic dark-energy-like fluid, and due to the NEC being satisfied, exotic matter is not required in this case.
\par We explored a different range of EoS parameters, $-1\leq\varpi_2\leq0.2$, to identify regions where the NEC is satisfied. The findings, summarized in Table~\ref{tab1}, indicate that for positive values of $\varpi_2$, traversable wormholes satisfying the NEC can exist-unlike in the Riemannian framework \cite{Tripathy}. In the Riemannian case, non-exotic wormhole solutions cease to exist for $\varpi_2\lesssim-0.36$, whereas the Finslerian geometry allows such solutions even in this domain.
\begin{table}[htpb]
\centering
\scriptsize
\setlength{\tabcolsep}{6.5pt}
\renewcommand{\arraystretch}{1.2}
\caption{The parameter range of $\varpi_2$ that ensures the satisfaction of the NEC for the exponential shape function at $r_0=0.5$ is identified.\label{tab1}}
{\begin{tabular}{@{}lccccclccccc@{}}
\toprule
${\textit{\textbf{\textrm{Case 1}}}}$ & $\xi$ & $\rho$ & $\rho+p_r$ & $\rho+p_t$ & $\triangle p$
& ${\textit{\textbf{\textrm{Case 1}}}}$ & $\xi$ & $\rho$ & $\rho+p_r$ & $\rho+p_t$ & $\triangle p$\\
 \hline
$\varpi_2=0.195$   & $\xi<-1$ & $+$ & Positive & Positive  &$+$   & $\varpi_2=0.175$   & $\xi<-1$ & $+$ & Positive & Positive  &$+$   \\ 
$\varpi_2=0.155$   & $\xi<-1$ & $+$ & Positive & Positive  &$+$   & $\varpi_2=0.12$    & $\xi<-1$ & $+$ & Positive & Positive  &$+$   \\
$\varpi_2=0.105$   & $\xi>-1$ & $+$ & Positive & Positive  &$+$   & $\varpi_2=0.10$    & $\xi<-1$ & $+$ & Positive & Positive  &$+$   \\ 
$\varpi_2=-0.13$   & $\xi>-1$ & $+$ & Positive & Positive  &$+$   & $\varpi_2=-0.15$   & $\xi>-1$ & $+$ & Positive & Positive  &$+$   \\
$\varpi_2=-0.165$  & $\xi<-1$ & $+$ & Positive & Positive  &$+$   &
$\varpi_2=-0.17$   & $\xi>-1$ & $+$ & Positive & Positive  &$+$   \\  
$\varpi_2=-0.18$   & $\xi>-1$ & $+$ & Positive & Positive  &$+$   &
$\varpi_2=-0.2$    & $\xi>-1$ & $+$ & Positive & Positive  &$-$   \\
$\varpi_2=-0.215$  & $\xi<-1$ & $+$ & Positive & Positive  &$-$   &
$\varpi_2=-0.22$   & $\xi>-1$ & $+$ & Positive & Positive  &$-$   \\
$\varpi_2=-0.24$   & $\xi<-1$ & $+$ & Positive & Positive  &$-$   &
$\varpi_2=-0.25$   & $\xi>-1$ & $+$ & Positive & Positive  &$-$   \\
$\varpi_2=-0.255$  & $\xi>-1$ & $+$ & Positive & Positive  &$-$   &
$\varpi_2=-0.26$   & $\xi<-1$ & $+$ & Positive & Positive  &$-$   \\
$\varpi_2=-0.265$  & $\xi<-1$ & $+$ & Positive & Positive  &$-$   &
$\varpi_2=-0.28$   & $\xi>-1$ & $+$ & Positive & Positive  &$-$   \\
$\varpi_2=-0.295$  & $\xi<-1$ & $+$ & Positive & Positive  &$-$   &
$\varpi_2=-0.305$  & $\xi<-1$ & $+$ & Positive & Positive  &$-$   \\
$\varpi_2=-0.315$  & $\xi<-1$ & $+$ & Positive & Positive  &$-$   &
$\varpi_2=-0.335$  & $\xi<-1$ & $+$ & Positive & Positive  &$-$   \\
$\varpi_2=-0.34$   & $\xi<-1$ & $+$ & Positive & Positive  &$-$   &
$\varpi_2=-0.355$  & $\xi>-1$ & $+$ & Positive & Positive  &$-$   \\
$\varpi_2=-0.365$  & $\xi<-1$ & $+$ & Positive & Positive  &$-$   &
$\varpi_2=-0.37$   & $\xi<-1$ & $+$ & Positive & Positive  &$-$   \\
$\varpi_2=-0.38$   & $\xi>-1$ & $+$ & Positive & Positive  &$-$   &
$\varpi_2=-0.39$   & $\xi<-1$ & $+$ & Positive & Positive  &$-$   \\
$\varpi_2=-0.43$   & $\xi<-1$ & $+$ & Positive & Positive  &$-$   &
$\varpi_2=-0.435$  & $\xi<-1$ & $+$ & Positive & Positive  &$-$   \\
$\varpi_2=-0.44$   & $\xi>-1$ & $+$ & Positive & Positive  &$-$   &
$\varpi_2=-0.445$  & $\xi>-1$ & $+$ & Positive & Positive  &$-$   \\
$\varpi_2=-0.45$   & $\xi>-1$ & $+$ & Positive & Positive  &$-$   &
$\varpi_2=-0.455$  & $\xi>-1$ & $+$ & Positive & Positive  &$-$   \\
$\varpi_2=-0.465$  & $\xi<-1$ & $+$ & Positive & Positive  &$-$   &
$\varpi_2=-0.58, ~~~ r>2.6$   & $\xi>-1$ & $+$ & Positive & Positive  &$-$ \\
$\varpi_2=-0.62, ~~~r>2.2$    & $\xi>-1$ & $+$ & Positive & Positive  &$-$   &
$\varpi_2=-0.655, ~~r>1.9$   & $\xi>-1$ & $+$ & Positive & Positive  &$-$   \\
$\varpi_2=-0.685, ~~r>1.8$   & $\xi>-1$ & $+$ & Positive & Positive  &$-$   &
$\varpi_2=-0.73, ~~~ r>1.6$   & $\xi<-1$ & $+$ & Positive & Positive  &$-$   \\
$\varpi_2=-0.815, ~~ r>1.4$   & $\xi<-1$ & $+$ & Positive & Positive  &$-$   &
$\varpi_2=-0.825, ~~ r>1.4$   & $\xi<-1$ & $+$ & Positive & Positive  &$-$   \\
$\varpi_2=-0.84, ~~~ r>1.3$   & $\xi>-1$ & $+$ & Positive & Positive  &$-$   &
$\varpi_2=-0.845, ~~ r>1.3$   & $\xi<-1$ & $+$ & Positive & Positive  &$-$   \\
$\varpi_2=-0.855, ~~ r>1.3$   & $\xi>-1$ & $+$ & Positive & Positive  &$-$   &
$\varpi_2=-0.86, ~~~ r>1.3$   & $\xi>-1$ & $+$ & Positive & Positive  &$-$   \\
$\varpi_2=-0.865, ~~ r>1.3$   & $\xi<-1$ & $+$ & Positive & Positive  &$-$   &
$\varpi_2=-0.87, ~~~ r>1.3$   & $\xi<-1$ & $+$ & Positive & Positive  &$-$   \\
$\varpi_2=-0.905, ~~ r>1.2$   & $\xi>-1$ & $+$ & Positive & Positive  &$-$   &
$\varpi_2=-0.915, ~~ r>1.2$   & $\xi>-1$ & $+$ & Positive & Positive  &$-$   \\
$\varpi_2=-0.925, ~~ r>1.2$   & $\xi>-1$ & $+$ & Positive & Positive  &$-$   &
$\varpi_2=-0.94, ~~~ r>1.2$   & $\xi<-1$ & $+$ & Positive & Positive  &$-$   \\
$\varpi_2=-0.945, ~~ r>1.2$   & $\xi>-1$ & $+$ & Positive & Positive  &$-$   &
$\varpi_2=-0.965, ~~ r>1.2$   & $\xi>-1$ & $+$ & Positive & Positive  &$-$   \\
$\varpi_2=-0.97, ~~~ r>1.2$   & $\xi<-1$ & $+$ & Positive & Positive  &$-$   &
$\varpi_2=-0.975, ~~ r>1.2$   & $\xi<-1$ & $+$ & Positive & Positive  &$-$   \\
$\varpi_2=-0.985, ~~ r>1.1$   & $\xi>-1$ & $+$ & Positive & Positive  &$-$   &
$\varpi_2=-0.995, ~~ r>1.1$   & $\xi>-1$ & $+$ & Positive & Positive  &$-$ \\
\hline 
 \end{tabular}}
 \end{table}
\subsection{Case 2:}
\begin{figure*}[htpb]
\centering
\mbox{{\includegraphics[scale=0.26]{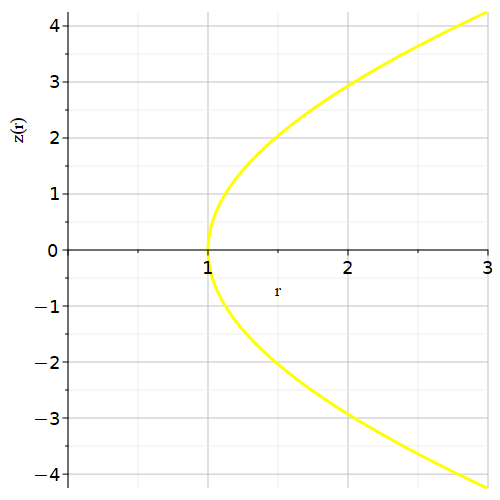}}
{\includegraphics[scale=0.35]{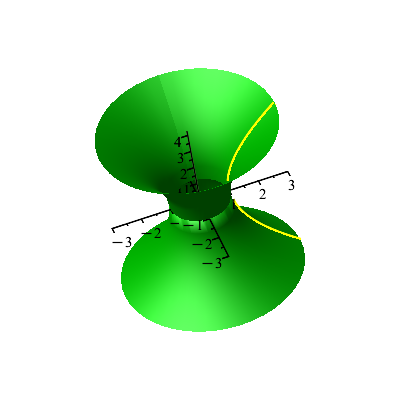}}{\includegraphics[scale=0.26]{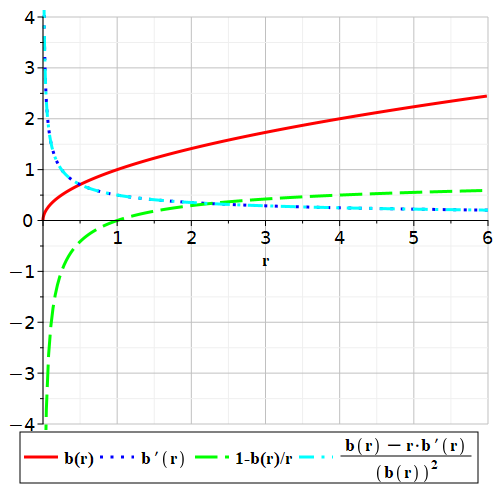}}}
\caption{\label{fig4} Embedding surfaces in two and three dimensions constructed from the power-law shape function $b(r)=\sqrt{r_0r}$ and its behavior at $r_0 = 1$.}
\end{figure*}

To model the wormhole geometry, we consider a power-law shape function given by $b(r)=\sqrt{r_0r}$, with its derivative $b^\prime(r)=\frac{\sqrt{r_0}}{2\sqrt{r}}$, leading to $\frac{b^\prime(r)}{b(r)}=\frac{1}{2r}$ \cite{Elizalde}. Figure~\ref{fig4} displays the shape function, confirming that it adheres to the required conditions.\\
Corresponding to this specific shape function, the resulting expressions for energy density, radial pressure, and tangential pressure are found to be,
\begin{eqnarray}
\rho&=&\frac{-1}{2(\xi+1)}\left[\frac{27+13\varpi_1+2\varpi_2}{(2\varpi_1\varpi_2+2\varpi_2+\varpi_1^2+3)(8\pi+3\ell_1)-12\ell_1+32\pi\varpi_1}\right]\sqrt{\frac{r_0}{r^5}},\label{Eq34}\\
p_r&=&\frac{2\varpi_2+\gamma}{2(\xi+1)}\left[\frac{27+13\varpi_1+2\varpi_2}{(2\varpi_1\varpi_2+2\varpi_2+\varpi_1^2+3)(8\pi+3\ell_1)-12\ell_1+32\pi\varpi_1}\right]\sqrt{\frac{r_0}{r^5}},\label{Eq35}\\
p_t&=&\frac{-\varpi_2}{2(\xi+1)}\left[\frac{27+13\varpi_1+2\varpi_2}{(2\varpi_1\varpi_2+2\varpi_2+\varpi_1^2+3)(8\pi+3\ell_1)-12\ell_1+32\pi\varpi_1}\right]\sqrt{\frac{r_0}{r^5}}.\label{Eq36}
\end{eqnarray}
In this case, we aim to understand whether traversable wormholes can exist without exotic matter, within the context of extended gravity theories applied in Finsler geometry. One of the studied configurations involves setting $\varpi_2=-0.465$, which reflects tangential tension similar to that of quintessence, resulting in $\varpi_1=0.33$ and indicating mild radial pressure. Under the condition $\xi<-1$, the energy density remains positive. We now investigate the corresponding energy conditions for a wormhole using a power-law shape function similar to case 1.
\begin{eqnarray}
\rho+p_r&=&\frac{1}{2(\xi+1)}\left[\frac{(2\varpi_2+\gamma-1)(27+13\varpi_1+2\varpi_2)}{(2\varpi_1\varpi_2+2\varpi_2+\varpi_1^2+3)(8\pi+3\ell_1)-12\ell_1+32\pi\varpi_1}\right]\sqrt{\frac{r_0}{r^5}},\label{Eq37}\\
\rho+p_t&=&\frac{1+\varpi_2}{2(\xi+1)}\left[\frac{-(27+13\varpi_1+2\varpi_2)}{(2\varpi_1\varpi_2+2\varpi_2+\varpi_1^2+3)(8\pi+3\ell_1)-12\ell_1+32\pi\varpi_1}\right]\sqrt{\frac{r_0}{r^5}},\label{Eq38}\\
\rho+p_r+2p_t&=&\frac{\gamma-1}{2(\xi+1)}\left[\frac{27+13\varpi_1+2\varpi_2}{(2\varpi_1\varpi_2+2\varpi_2+\varpi_1^2+3)(8\pi+3\ell_1)-12\ell_1+32\pi\varpi_1}\right]\sqrt{\frac{r_0}{r^5}},\label{Eq39}\\
\rho-p_r&=&\frac{1}{2(\xi+1)}\left[\frac{-(2\varpi_2+\gamma+1)(27+13\varpi_1+2\varpi_2)}{(2\varpi_1\varpi_2+2\varpi_2+\varpi_1^2+3)(8\pi+3\ell_1)-12\ell_1+32\pi\varpi_1}\right]\sqrt{\frac{r_0}{r^5}},\label{Eq40}\\
\rho-p_t&=&\frac{\varpi_2-1}{2(\xi+1)}\left[\frac{27+13\varpi_1+2\varpi_2}{(2\varpi_1\varpi_2+2\varpi_2+\varpi_1^2+3)(8\pi+3\ell_1)-12\ell_1+32\pi\varpi_1}\right]\sqrt{\frac{r_0}{r^5}}.\label{Eq41}
\end{eqnarray}
\par
\begin{figure*}[htpb]
\centering
\mbox{{\includegraphics[scale=0.26]{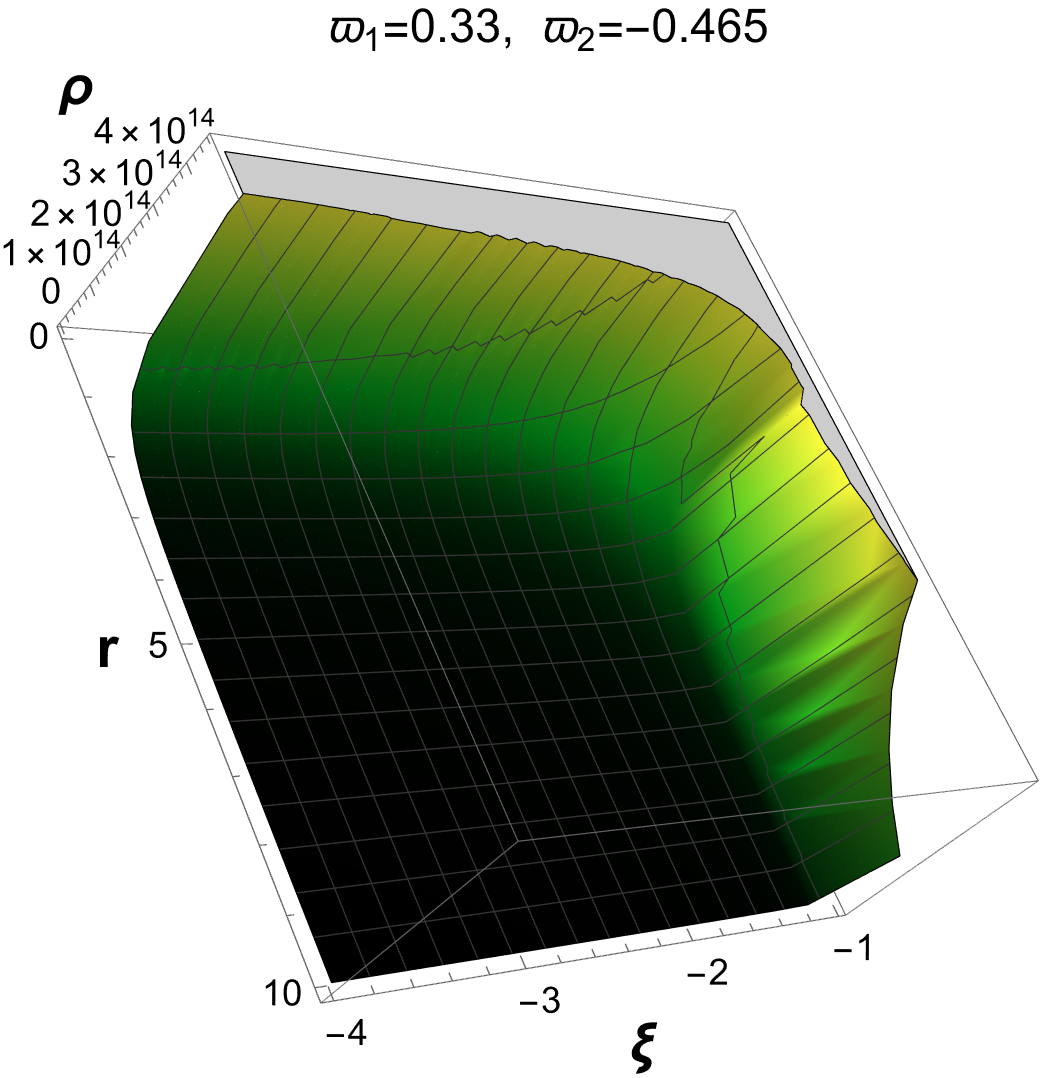}}
{\includegraphics[scale=0.26]{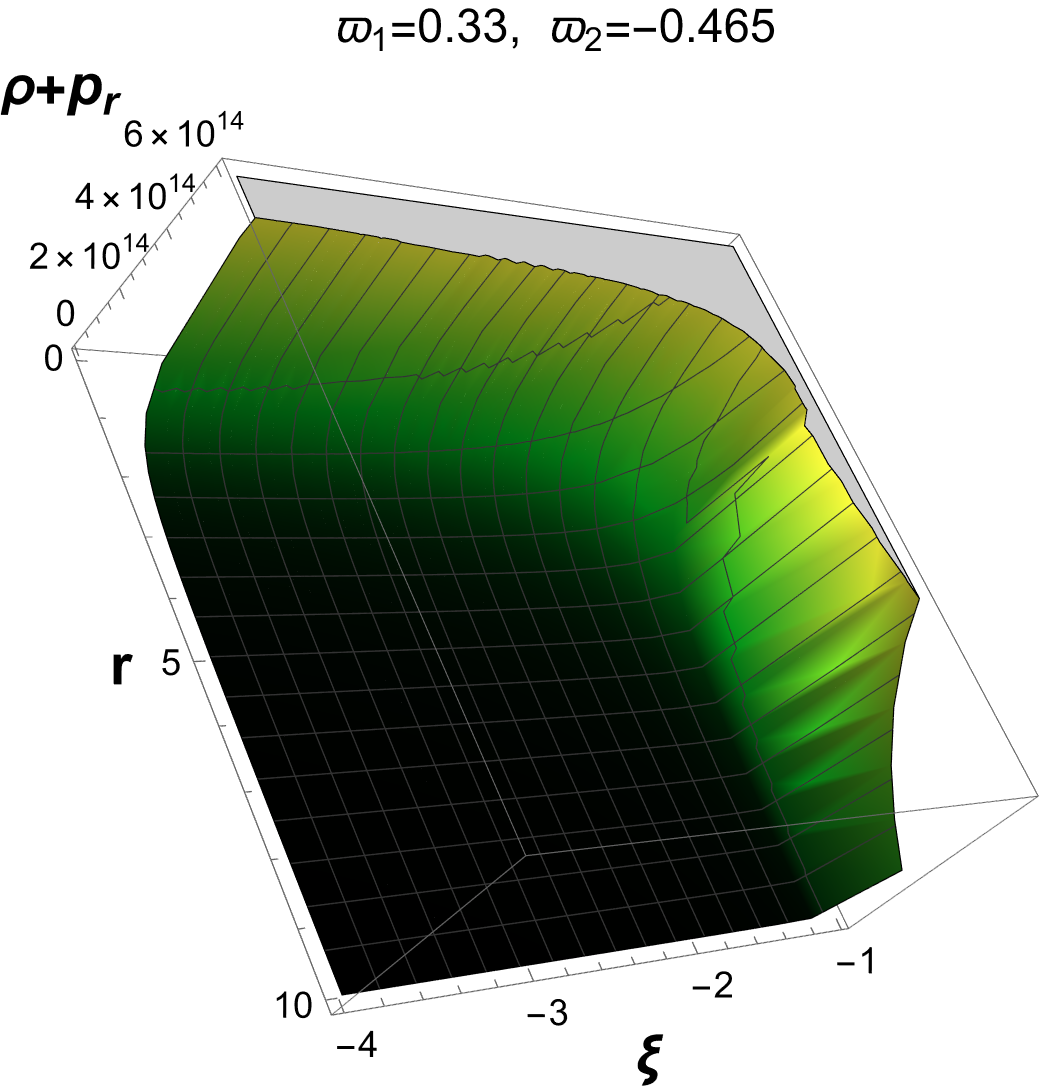}}{\includegraphics[scale=0.26]{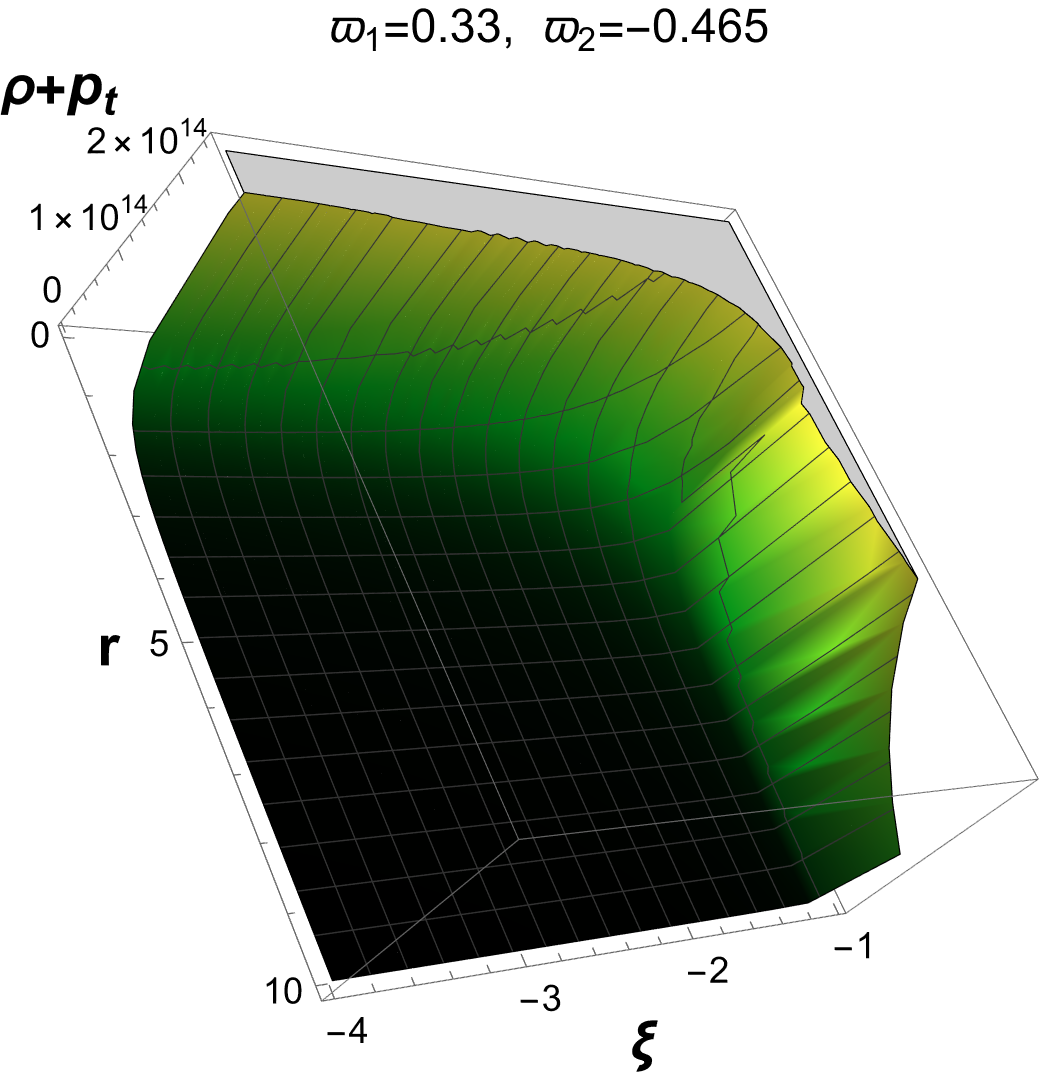}}}\\
\mbox{{\includegraphics[scale=0.3]{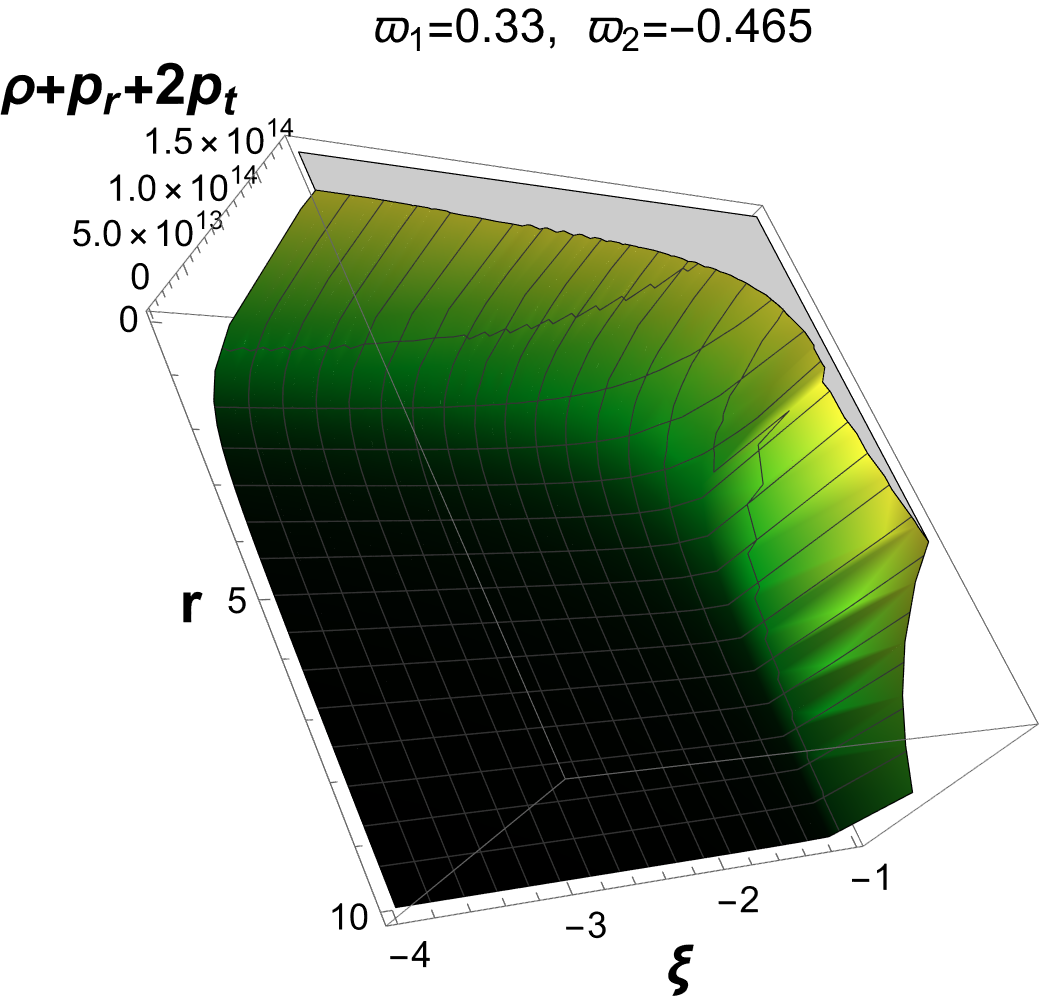}}
{\includegraphics[scale=0.26]{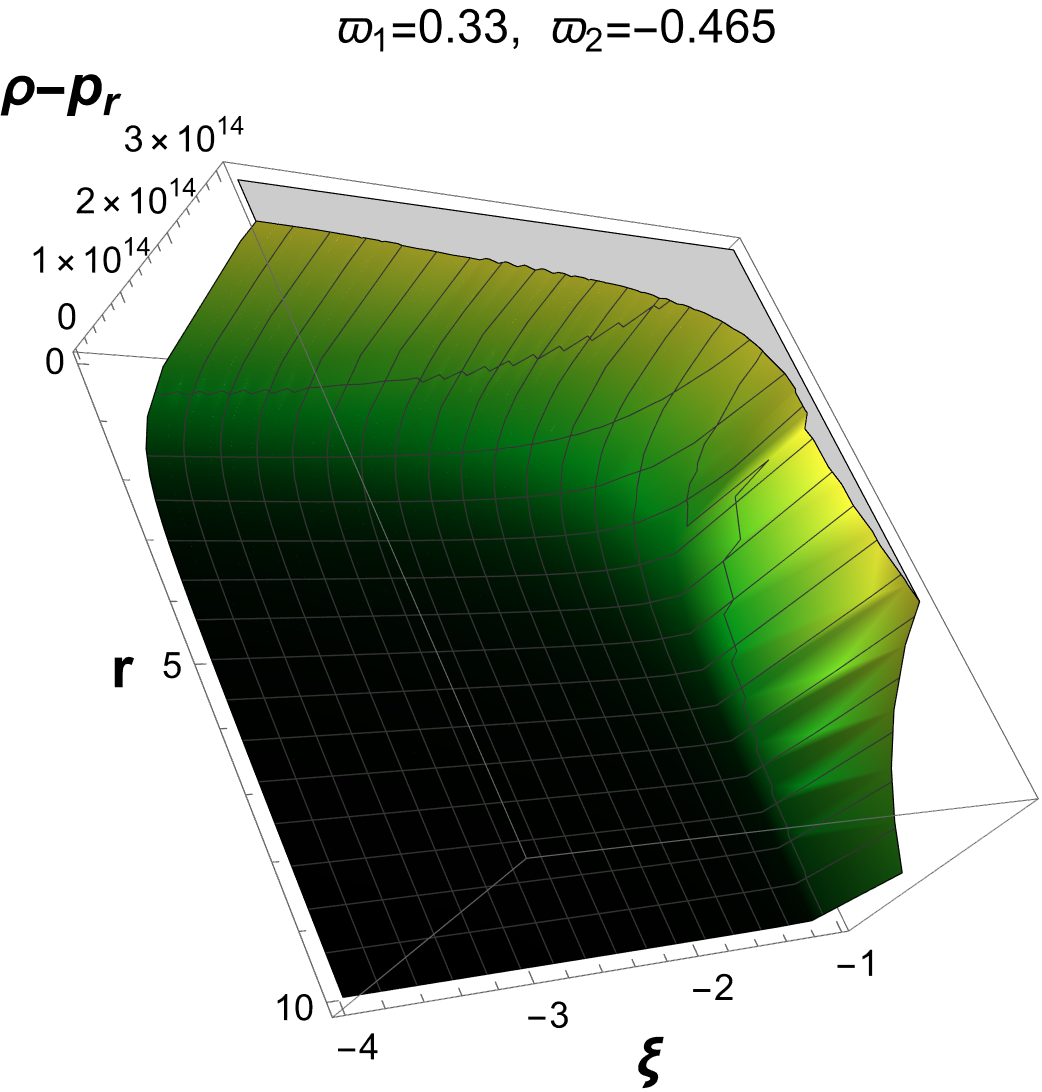}}{\includegraphics[scale=0.26]{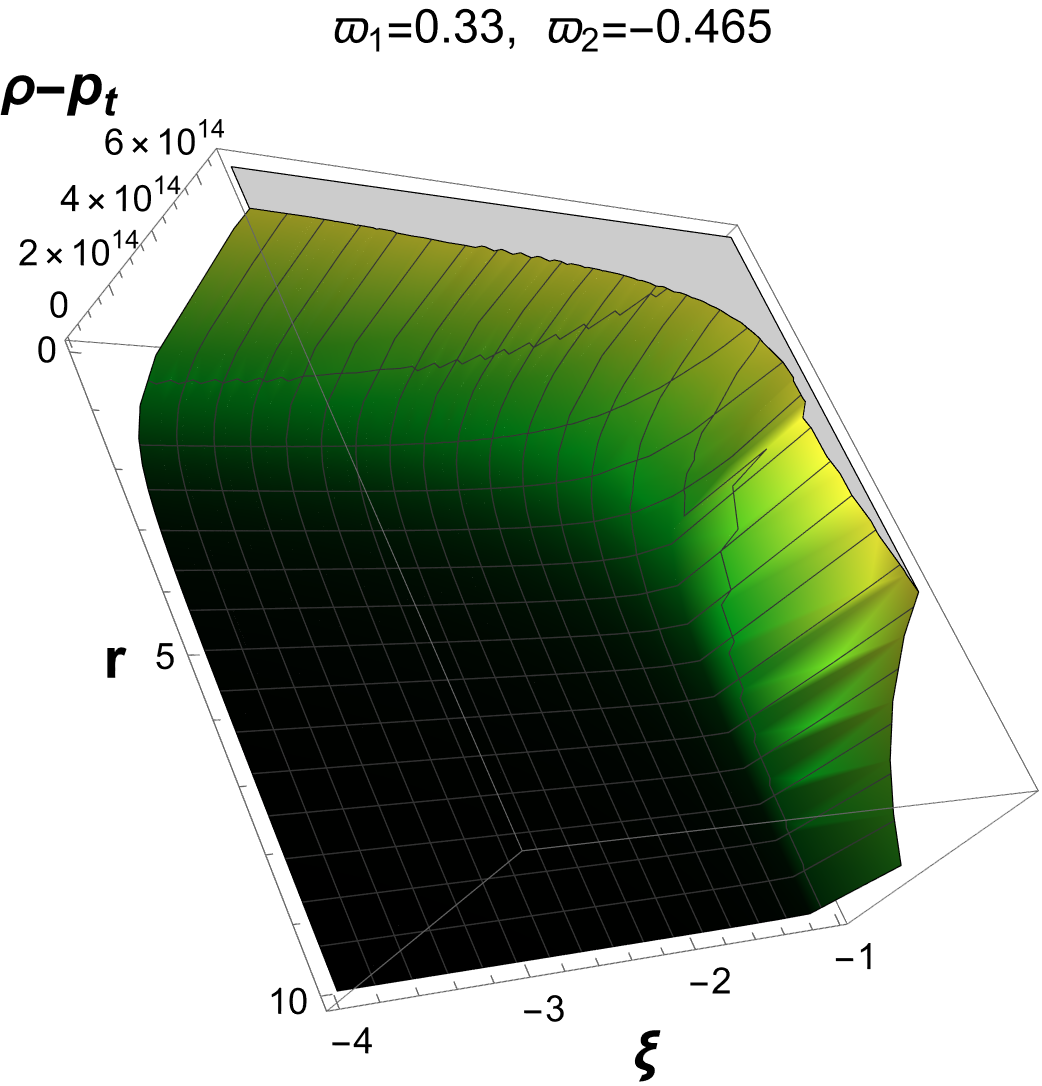}}}
\caption{\label{fig5} Satisfaction of all energy conditions (NEC, WEC, SED, DEC) at $r_0=1$ and $\xi<-1$.}
\end{figure*}

 Figure~\ref{fig5} illustrates all the energy conditions, indicating that for the selected value of $\varpi_2$, the wormhole matter content satisfies the NEC, thus supporting the existence of non-exotic matter wormholes.\\
We further explored various values of the EoS parameter to identify the specific range for which NEC remains valid. These findings are presented in Table~\ref{tab2}. In contrast to the Riemannian scenario \cite{Tripathy}, NEC is satisfied within the range $-0.36\leq\varpi_2\leq0.12$. For $\varpi_2<-1$, non-exotic matter wormholes are not achievable within this framework of geometry and gravity theory.\\ 
\begin{table}[ht]
\caption{The parameter range of $\varpi_2$ that ensures the satisfaction of the NEC for the power-law shape function at $r_0=1$ is identified.\label{tab2}}
{\begin{tabular}{@{}lccccclccccc@{}}
\toprule
${\textit{\textbf{\textrm{Case 2}}}}$ & $\xi$ & $\rho$ & $\rho+p_r$ & $\rho+p_t$ & $\triangle p$&
${\textit{\textbf{\textrm{Case 2}}}}$ & $\xi$ & $\rho$ & $\rho+p_r$ & $\rho+p_t$ & $\triangle p$\\
 \hline
$\varpi_2=0.195$   & $\xi<-1$ & $+$ & Positive & Positive  &$+$   &
$\varpi_2=0.175$   & $\xi<-1$ & $+$ & Positive & Positive  &$+$   \\
$\varpi_2=0.155$   & $\xi<-1$ & $+$ & Positive & Positive  &$+$   &   
$\varpi_2=0.12$    & $\xi<-1$ & $+$ & Positive & Positive  &$+$   \\
$\varpi_2=0.105$   & $\xi>-1$ & $+$ & Positive & Positive  &$+$   &  
$\varpi_2=0.10$    & $\xi<-1$ & $+$ & Positive & Positive  &$+$   \\ 
$\varpi_2=-0.13$   & $\xi>-1$ & $+$ & Positive & Positive  &$+$   &
$\varpi_2=-0.15$   & $\xi>-1$ & $+$ & Positive & Positive  &$+$   \\
$\varpi_2=-0.165$  & $\xi<-1$ & $+$ & Positive & Positive  &$+$   &
$\varpi_2=-0.17$   & $\xi>-1$ & $+$ & Positive & Positive  &$+$   \\  
$\varpi_2=-0.18$   & $\xi>-1$ & $+$ & Positive & Positive  &$+$   &
$\varpi_2=-0.2$    & $\xi>-1$ & $+$ & Positive & Positive  &$-$   \\
$\varpi_2=-0.215$  & $\xi<-1$ & $+$ & Positive & Positive  &$-$   & 
$\varpi_2=-0.22$   & $\xi>-1$ & $+$ & Positive & Positive  &$-$   \\
$\varpi_2=-0.24$   & $\xi<-1$ & $+$ & Positive & Positive  &$-$   &
$\varpi_2=-0.25$   & $\xi>-1$ & $+$ & Positive & Positive  &$-$   \\
$\varpi_2=-0.255$  & $\xi>-1$ & $+$ & Positive & Positive  &$-$   & 
$\varpi_2=-0.26$   & $\xi<-1$ & $+$ & Positive & Positive  &$-$   \\
$\varpi_2=-0.265$  & $\xi<-1$ & $+$ & Positive & Positive  &$-$   & 
$\varpi_2=-0.28$   & $\xi>-1$ & $+$ & Positive & Positive  &$-$   \\
$\varpi_2=-0.295$  & $\xi<-1$ & $+$ & Positive & Positive  &$-$   &
$\varpi_2=-0.305$  & $\xi<-1$ & $+$ & Positive & Positive  &$-$   \\ 
$\varpi_2=-0.315$  & $\xi<-1$ & $+$ & Positive & Positive  &$-$   & 
$\varpi_2=-0.335$  & $\xi<-1$ & $+$ & Positive & Positive  &$-$   \\ 
$\varpi_2=-0.34$   & $\xi<-1$ & $+$ & Positive & Positive  &$-$   &
$\varpi_2=-0.355$  & $\xi>-1$ & $+$ & Positive & Positive  &$-$   \\
$\varpi_2=-0.365$  & $\xi<-1$ & $+$ & Positive & Positive  &$-$   &  
$\varpi_2=-0.37$   & $\xi<-1$ & $+$ & Positive & Positive  &$-$   \\
$\varpi_2=-0.38$   & $\xi>-1$ & $+$ & Positive & Positive  &$-$   &
$\varpi_2=-0.39$   & $\xi<-1$ & $+$ & Positive & Positive  &$-$   \\
$\varpi_2=-0.43$   & $\xi<-1$ & $+$ & Positive & Positive  &$-$   &
$\varpi_2=-0.435$  & $\xi<-1$ & $+$ & Positive & Positive  &$-$   \\ 
$\varpi_2=-0.44$   & $\xi>-1$ & $+$ & Positive & Positive  &$-$   &
$\varpi_2=-0.445$  & $\xi>-1$ & $+$ & Positive & Positive  &$-$   \\ 
$\varpi_2=-0.45$   & $\xi>-1$ & $+$ & Positive & Positive  &$-$   &
$\varpi_2=-0.455$  & $\xi>-1$ & $+$ & Positive & Positive  &$-$   \\
$\varpi_2=-0.465$  & $\xi<-1$ & $+$ & Positive & Positive  &$-$   &
$\varpi_2=-0.58$   & $\xi<-1$ & $+$ & Positive & Positive  &$-$   \\
$\varpi_2=-0.62$   & $\xi<-1$ & $+$ & Positive & Positive  &$-$   &
$\varpi_2=-0.655$  & $\xi<-1$ & $+$ & Positive & Positive  &$-$   \\
$\varpi_2=-0.685$  & $\xi<-1$ & $+$ & Positive & Positive  &$-$   &
$\varpi_2=-0.73$   & $\xi>-1$ & $+$ & Positive & Positive  &$-$   \\
$\varpi_2=-0.815$  & $\xi>-1$ & $+$ & Positive & Positive  &$-$   &
$\varpi_2=-0.825$  & $\xi>-1$ & $+$ & Positive & Positive  &$-$   \\
$\varpi_2=-0.84$   & $\xi<-1$ & $+$ & Positive & Positive  &$-$   &
$\varpi_2=-0.845$  & $\xi>-1$ & $+$ & Positive & Positive  &$-$   \\
$\varpi_2=-0.855$  & $\xi<-1$ & $+$ & Positive & Positive  &$-$   &
$\varpi_2=-0.86$   & $\xi<-1$ & $+$ & Positive & Positive  &$-$   \\
$\varpi_2=-0.865$  & $\xi>-1$ & $+$ & Positive & Positive  &$-$   &
$\varpi_2=-0.87$   & $\xi>-1$ & $+$ & Positive & Positive  &$-$   \\
$\varpi_2=-0.905$  & $\xi<-1$ & $+$ & Positive & Positive  &$-$   &
$\varpi_2=-0.915$  & $\xi<-1$ & $+$ & Positive & Positive  &$-$   \\
$\varpi_2=-0.925$  & $\xi<-1$ & $+$ & Positive & Positive  &$-$   &
$\varpi_2=-0.94$   & $\xi>-1$ & $+$ & Positive & Positive  &$-$   \\
$\varpi_2=-0.945$  & $\xi<-1$ & $+$ & Positive & Positive  &$-$   &
$\varpi_2=-0.965$  & $\xi<-1$ & $+$ & Positive & Positive  &$-$   \\
$\varpi_2=-0.97$   & $\xi>-1$ & $+$ & Positive & Positive  &$-$   &
$\varpi_2=-0.975$  & $\xi>-1$ & $+$ & Positive & Positive  &$-$   \\
$\varpi_2=-0.985$  & $\xi<-1$ & $+$ & Positive & Positive  &$-$   &
$\varpi_2=-0.995$  & $\xi<-1$ & $+$ & Positive & Positive  &$-$   \\
\hline
 \end{tabular}}
\end{table}

\par It is noteworthy that, unlike the standard GR scenario, the energy conditions in the present model remain satisfied even in the presence of quintessence-like matter. This behavior can be understood by rewriting the modified field equations in terms of an effective energy–momentum tensor, where additional contributions arise due to the matter–geometry coupling in the squared trace $\mathfrak{f}(\mathds{R},\mathds{T})$ gravity. These extra terms effectively act as a gravitational source, allowing the wormhole geometry to satisfy the flaring-out condition without requiring violation of the null energy condition by the physical matter. Furthermore, the Finsler–Randers structure introduces intrinsic anisotropy, leading to different radial and tangential pressures. In particular, the negative tangential pressure (quintessence-like behavior) is balanced by the modified gravity contributions and the radial pressure, resulting in a configuration that satisfies $\rho+p_r\geq 0$ and $\rho+p_t\geq 0$. Therefore, the combined effect of modified gravity and Finslerian anisotropy shifts the requirement of exotic matter from the physical sector to the effective geometric sector, enabling the existence of non-exotic traversable wormhole solutions.
To establish a rigorous quantitative baseline comparison, we examine the parameter boundaries required to successfully satisfy the Null Energy Condition (NEC) at the wormhole throat radius. In conventional Riemannian frameworks ($\xi = 0$), non-exotic traversable wormholes are highly constrained, with physically viable configurations completely failing to exist once the tangential equation-of-state parameter falls below the critical threshold of $\varpi_2 \lesssim -0.36$. Conversely, generalizing the space-time configuration to a Finsler-Randers geometry ($\xi \neq 0$) notably shifts these physical boundaries for both the exponential and power-law shape functions. This directional geometric buffering allows the NEC to remain fully satisfied over a significantly expanded physical envelope that extends down to $\varpi_2 \approx -1.0$, which directly proves that the modified gravitational background relaxes the strict parameter fine-tuning required in standard general relativity.
\subsection{Anisotropy Factor}
\begin{figure*}[htpb]
\centering
\mbox{{\includegraphics[scale=0.31]{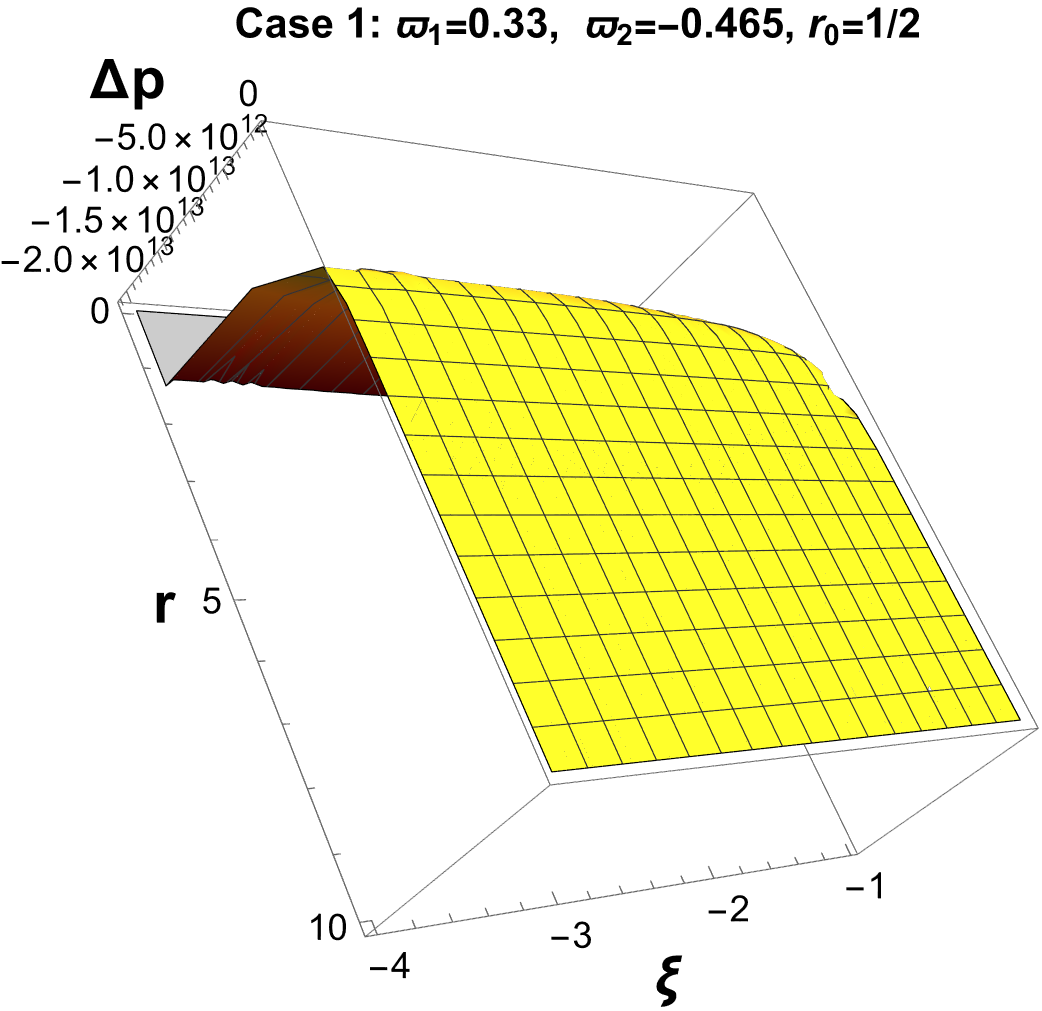}}{\includegraphics[scale=0.3]{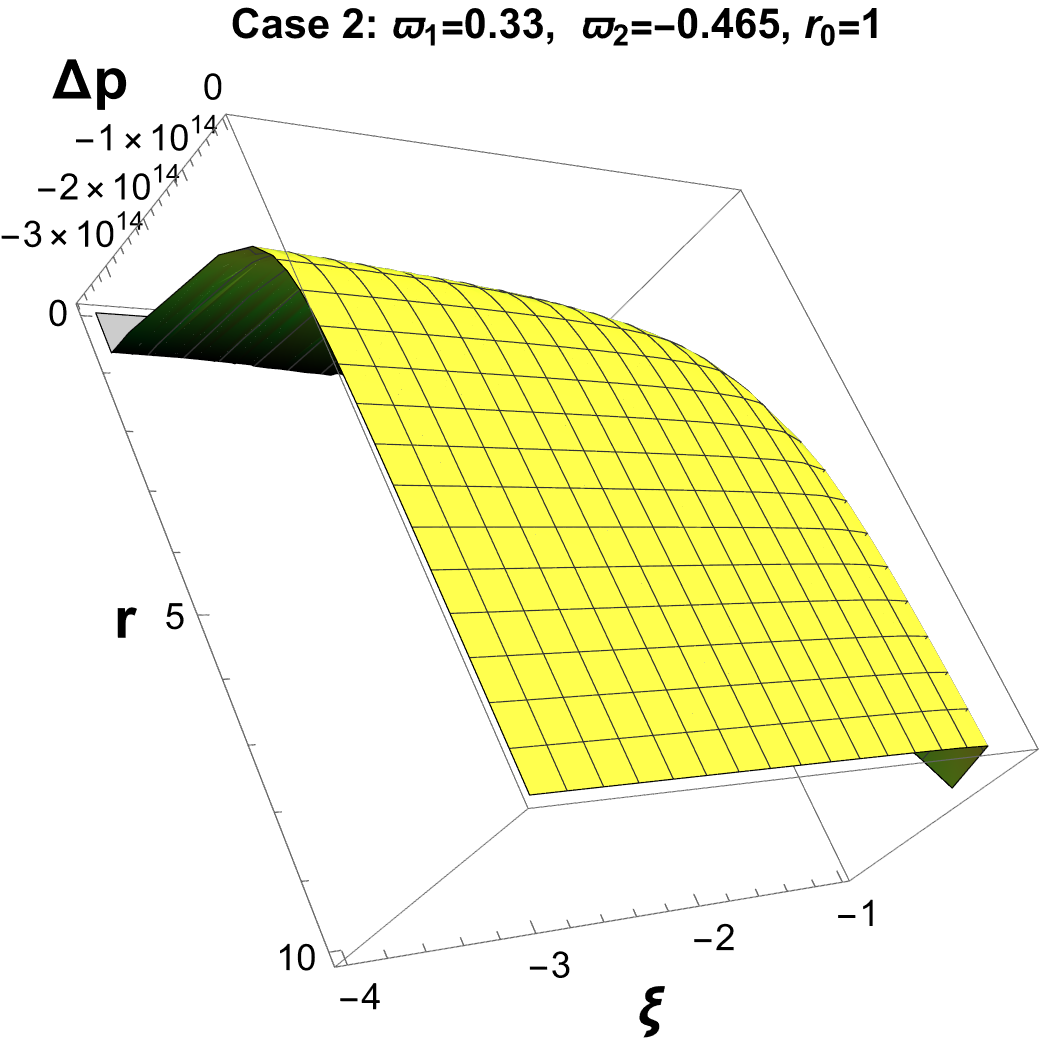}}}
\caption{\label{fig6} Depictions of the pressure anisotropy ($\triangle p$).}
\end{figure*}

Finsler geometry generalizes Riemannian geometry by introducing direction-dependent metrics, leading to more general Einstein-like field equations and effective matter sources. It naturally accommodates anisotropic fluids. Anisotropy exists when $p_r\neq p_t$.
$\triangle p=p_t-p_r$, known as the anisotropy factor \cite{Rahaman, Sharif}. The anisotropy in pressure for this wormhole configuration can be written as,
\begin{eqnarray}
\textrm{Case 1} &:& \triangle p=\left[\frac{-2r(3-11\varpi_1+2\varpi_2)-12(\varpi_1+1)}{(2\varpi_1\varpi_2+2\varpi_2+\varpi_1^2+3)(8\pi+3\ell_1)-12\ell_1
+32\pi\varpi_1}\right]\frac{3\varpi_2+\gamma}{(\xi+1)r^2\textrm{e}^{2(r-r_0)}},\label{Eq42}\\
\textrm{Case 2} &:& \triangle 
p=\frac{3\varpi_2+\gamma}{2(\xi+1)}\left[\frac{-(27+13\varpi_1+2\varpi_2)}{(2\varpi_1\varpi_2+2\varpi_2+\varpi_1^2+3)(8\pi+3\ell_1)-12\ell_1+32\pi\varpi_1}\right]\sqrt{\frac{r_0}{r^5}}.\label{Eq43}
\end{eqnarray}
From the graphical profile of $\triangle p$ in Fig.~\ref{fig6}, it can be inferred that the wormhole geometry in both cases tends toward an attractive. The tangential pressure is more negative than the radial pressure. This implies that there is stronger tension in the tangential directions, which helps keep the wormhole throat open. A negative $\triangle p$ is a common feature in traversable wormhole solutions- it supports the flaring-out condition configuration.\\
\subsection{Stability of Wormhole Solution}
The stability of the obtained wormhole solutions is investigated through the generalized Tolman–Oppenheimer–Volkoff (TOV) equation for anisotropic matter distribution, given by
\begin{equation}
-\frac{\Psi^\prime}{2}(\rho+p_r)-\frac{dp_r}{dr}+\frac{2}{r}(p_t-p_r)=0.
\end{equation}
The above equation represents the equilibrium condition for the wormhole configuration under the combined action of gravitational ($F_g$), hydrostatic ($F_h$), and anisotropic ($F_a$) forces, respectively, where
\begin{eqnarray}
F_g&=&-\frac{\Psi^\prime}{2}(\rho+p_r),\\
F_h&=&-\frac{dp_r}{dr},\\
F_a&=&\frac{2}{r}(p_t-p_r).
\end{eqnarray}
Since a constant redshift function is adopted in the present work, i.e., $\Psi^\prime(r)=0$, the gravitational force vanishes identically ($F_g=0$). Consequently, the equilibrium of the wormhole geometry is governed by the interplay between the hydrostatic and anisotropic forces.
\begin{figure*}[htpb]
\centering
\mbox{{\includegraphics[scale=0.32]{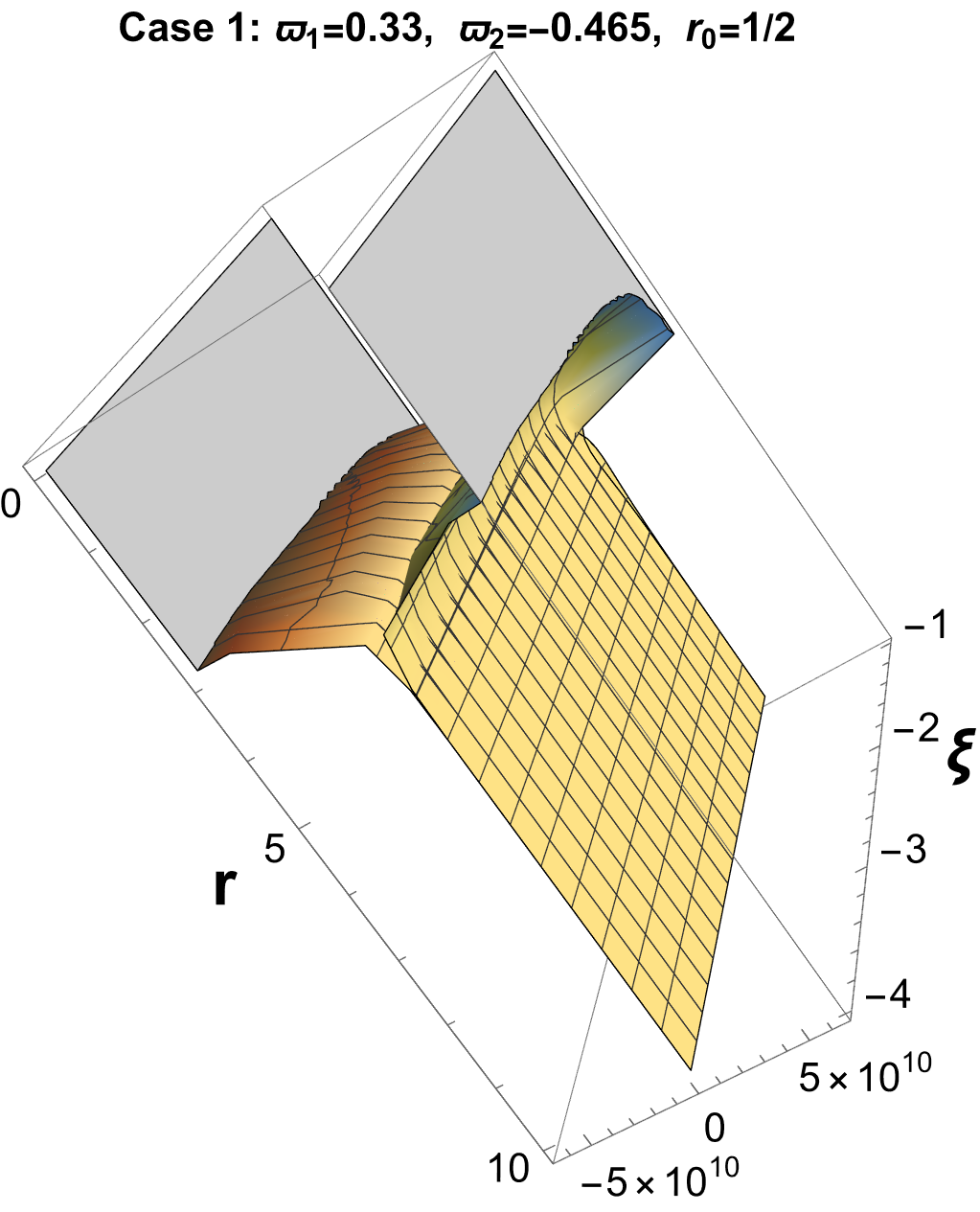}}{\includegraphics[scale=0.29]{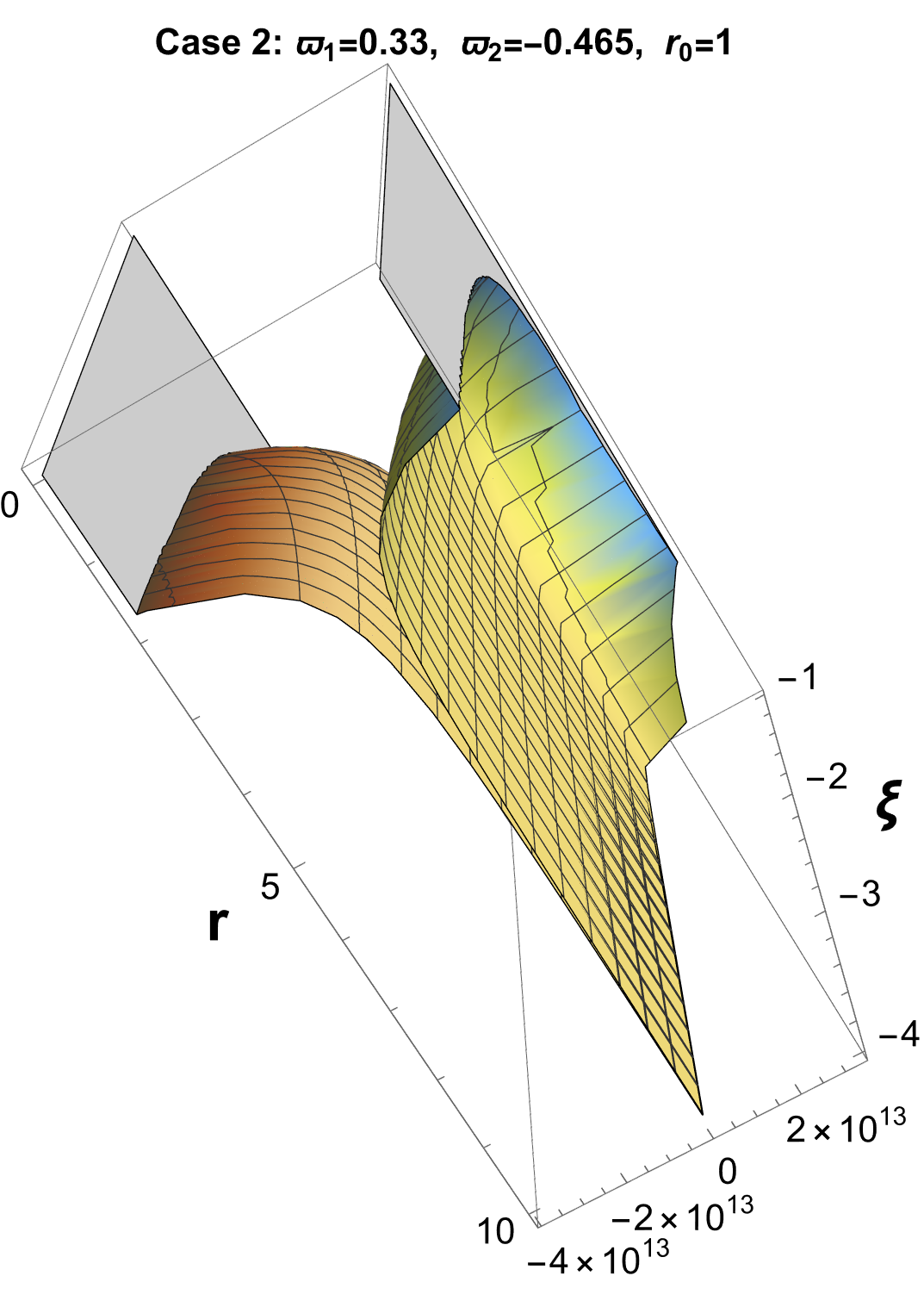}}}
\caption{\label{fig61} The graphical representation of hydrostatic force $F_h$ and anisotropic force $F_a$.}
\end{figure*}
The behavior of these forces is illustrated in Fig. \ref{fig61} for two cases. It is observed that the hydrostatic and anisotropic forces exhibit opposite tendencies throughout the considered domain, indicating that the anisotropic pressure effectively counterbalances the hydrostatic force. This qualitative force balance suggests that the obtained wormhole configurations remain in an equilibrium state and hence support physically viable traversable wormhole solutions within the present framework.\\

\section{A Geometric Measure of Exoticity in Traversable Wormholes}\label{sec6}
Though quantum effects may allow for NEC violations supporting wormhole throats \cite{Hochberg}, and such violations are common in GR, alternative gravity theories can permit wormholes sustained entirely by non-exotic matter. This investigation reveals that non-exotic matter wormholes are attainable for specific intervals of the EoS parameter, highlighting the importance of quantifying the exotic matter contribution. To evaluate the exotic matter contribution, we utilize the following formulation \cite{Tripathy1,Kar}
\begin{equation}\label{Eq44}
m=8\pi\oint(p+p_r)dV=8\pi\int_{r_0}^{\infty}(\rho+p_r)r^2dr.
\end{equation}
The radial NEC, $\rho+p_r\geq0$, is crucial at the wormhole throat $r=r_0$ because the flare-out condition (which ensures a wormhole rather than a black hole or regular geometry) is directly linked to this quantity. A violation of $\rho+p_r\geq0$ is what typically signals the presence of exotic matter needed to sustain the throat. This integral is valuable to assess whether exotic matter is required near the throat, not just pointwise, but as a net effect, taking the geometry into account. It's crucial for evaluating the physical viability of wormholes in alternative theories of gravity such as Finsler geometry.
$m$ can be interpreted as a measure of the effective energy supporting the wormhole, specifically: (1) The active gravitational mass contribution from $\rho+p_r$ over the spatial volume. (2) Sometimes called the "averaged energy condition integral", or a measure of how much non-exoticity (or exoticity) is present. 
\begin{itemize}
\item If $m<0$, indicating NEC violation, i.e., exotic matter is required somewhere in the wormhole structure.
\item If $m>0$, then non-exotic matter is sufficient to support the geometry (at least radially).
\end{itemize}
A full-range integral $\int_{r_0}^{\infty}(\rho+p_r)dV$ is less informative because (I) it mixes throat behavior with asymptotic regions. (II) Exotic matter, if required, is almost always localized near the throat. (III) Matter becomes normal (or vacuum-like) farther out.
\subsection{Case 1:}
\begin{figure*}[htpb]
\centering
{\includegraphics[scale=0.29]{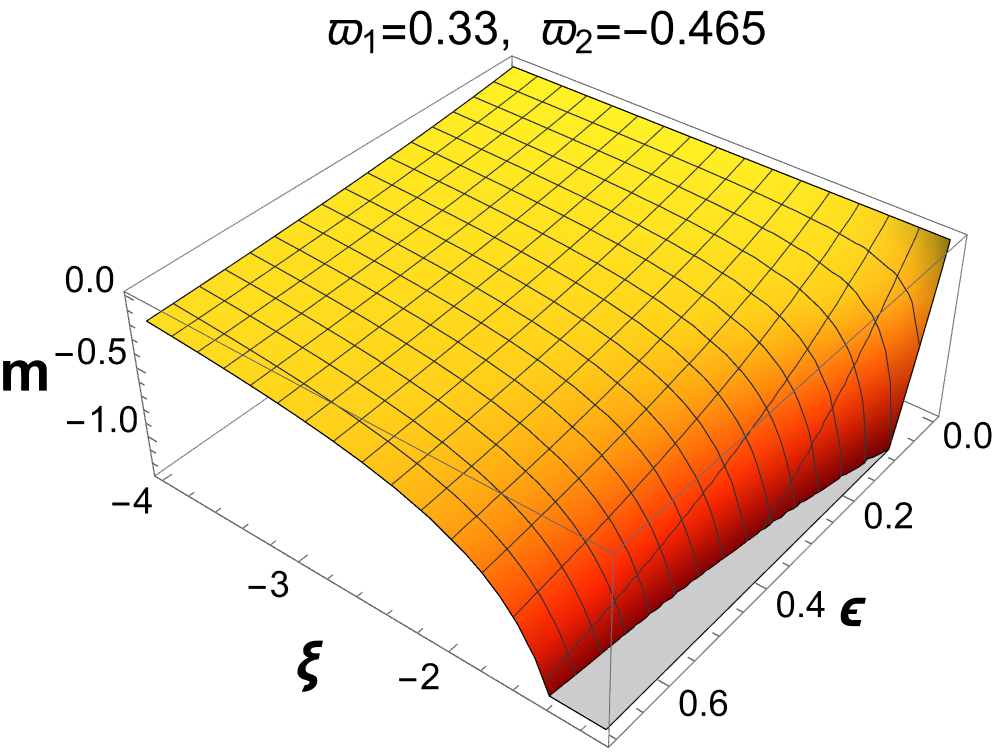}}\\\mbox{{\includegraphics[scale=0.32]{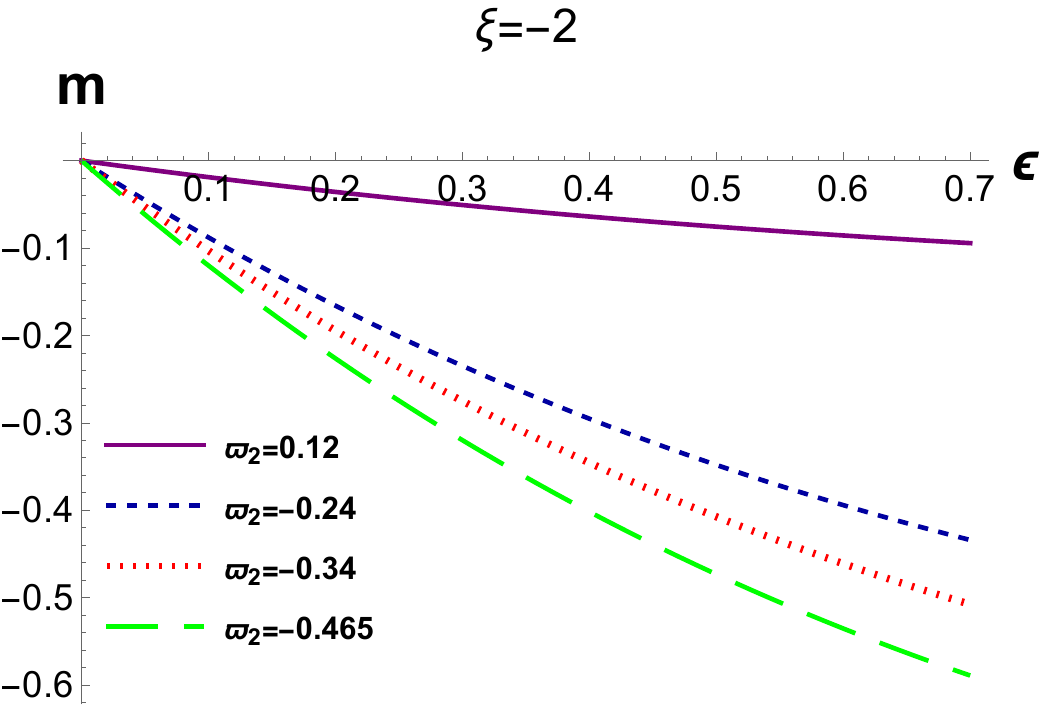}}{\includegraphics[scale=0.32]{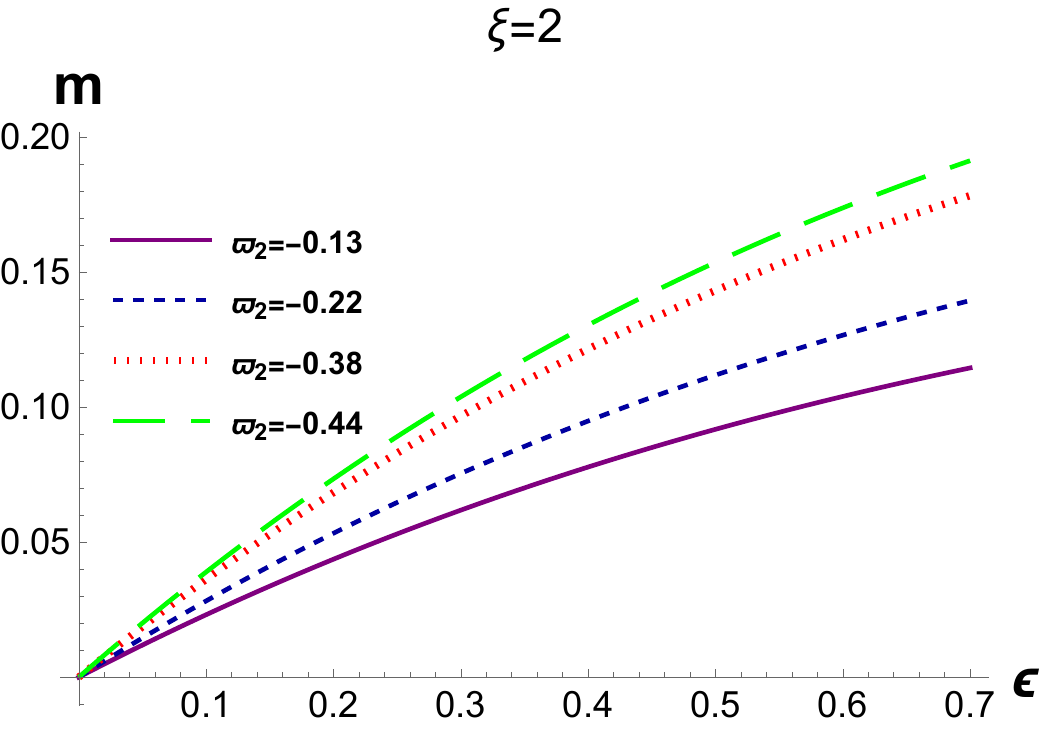}}}
\caption{\label{fig7} Exploring exotic matter content corresponding to specific $\varpi_2$ in the context of an exponential shape function with $r_0=1/2$.}
\end{figure*}

While $\rho+p_r>0$ might hold pointwise in most regions, small NEC-violating zones may still exist near the throat. The volume integral $\int_{r_0}^{r_0+\epsilon}(\rho+p_r)dV$ measures the net effect of such local violations. The result of the integral, when applied to the exponential shape function, is found to be,
\begin{eqnarray}
m&=&8\pi\int_{r_0}^{r_0+\epsilon}(\rho+p_r)r^2dr\nonumber\\\hspace{-2cm}&=&\frac{\bigg[\epsilon(-6+22\varpi_1-4\varpi_2)+(-1+\textrm{e}^{2\epsilon})(15+\varpi_1+2\varpi_2)
-2(-1+\textrm{e}^{2\epsilon})(-3+11\varpi_1-2\varpi_2)r_0\bigg]}{(\xi+1)(15-12\pi+8\varpi_1+5\varpi_1^2+10\varpi_2+5\varpi_1\varpi_2)}\nonumber\\&\times&\left[-\textrm{e}^{-2\epsilon}(1+\varpi_1)\right],\label{Eq45}
\end{eqnarray}
where $\epsilon<1$. Given a particular choice of the EoS parameter ($\varpi_2$), the calculation of exotic matter content becomes straightforward. In Fig.~\ref{fig7}, we present the exotic mass content for wormholes employing an exponential function, plotted against $\epsilon$.
Even though the matter content appears non-exotic locally (due to positive $\rho$ and small $p_r$), the Finslerian geometric effects (due to $\xi<-1$) near the throat creates a localized region where the volume-integrated radial NEC is violated - effectively behaving like exotic matter is present due to geometry, not due to the EoS. The two-dimensional plots in Fig.~\ref{fig7} are plotted for the specific values mentioned in Table~\ref{tab1}. The results show that the value of $m$ is positive for values of $\varpi_2$ that depend on $\xi>-1$. So both local and integrated radial NEC are satisfied. In the vicinity just beyond the throat radius, the wormhole exhibits non-exotic mass content.
\subsection{Case 2:}
\begin{figure*}[htpb]
\centering
{\includegraphics[scale=0.27]{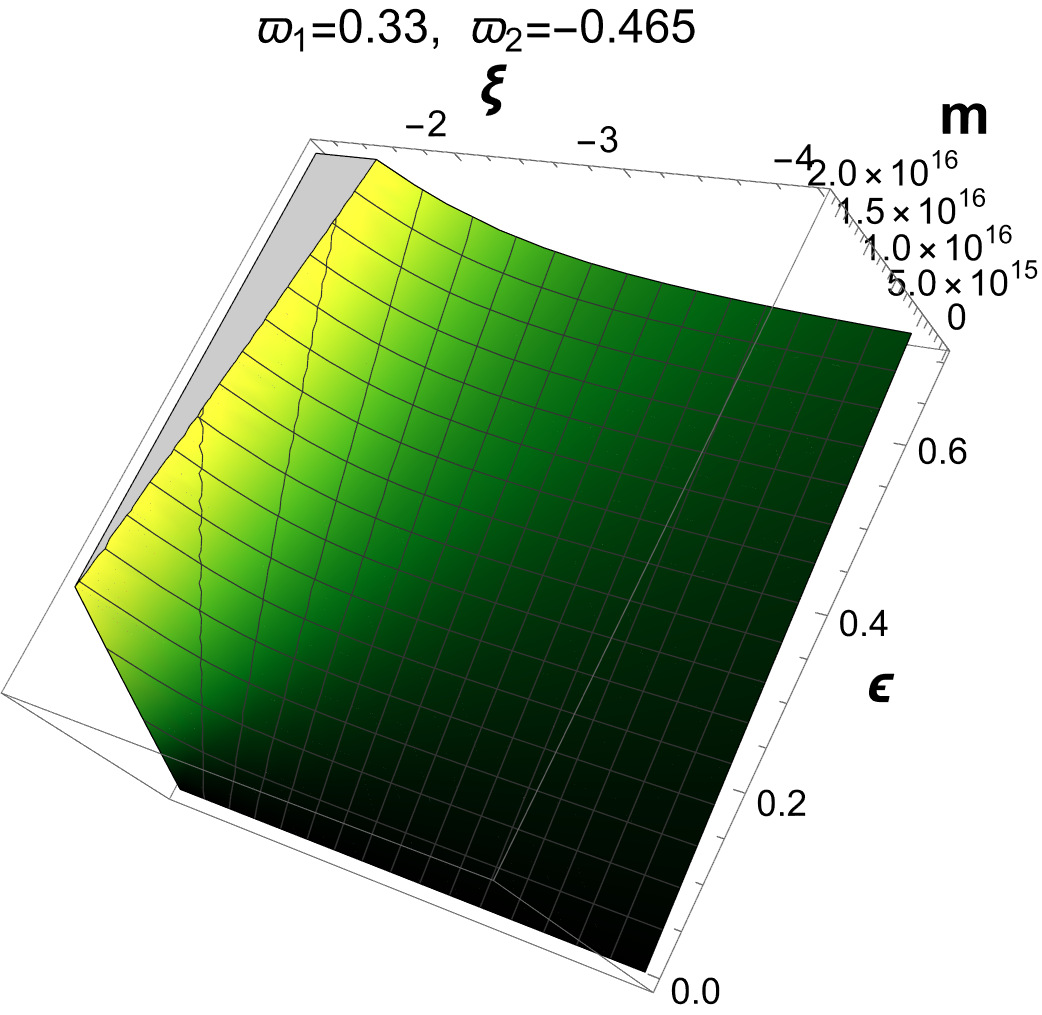}}\\\mbox{{\includegraphics[scale=0.32]{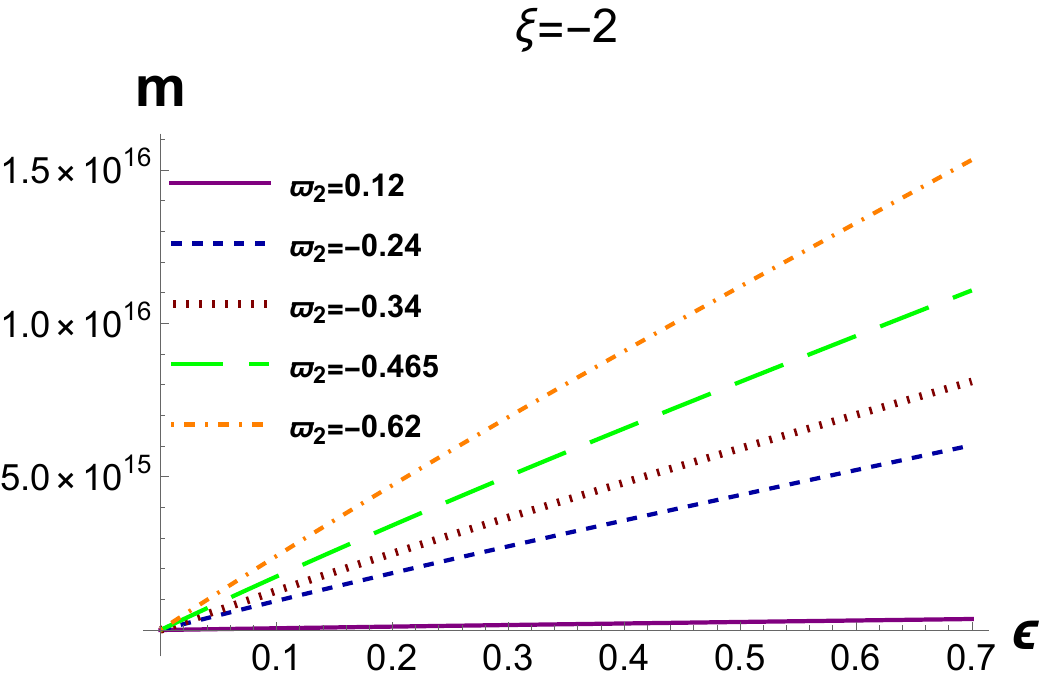}}{\includegraphics[scale=0.32]{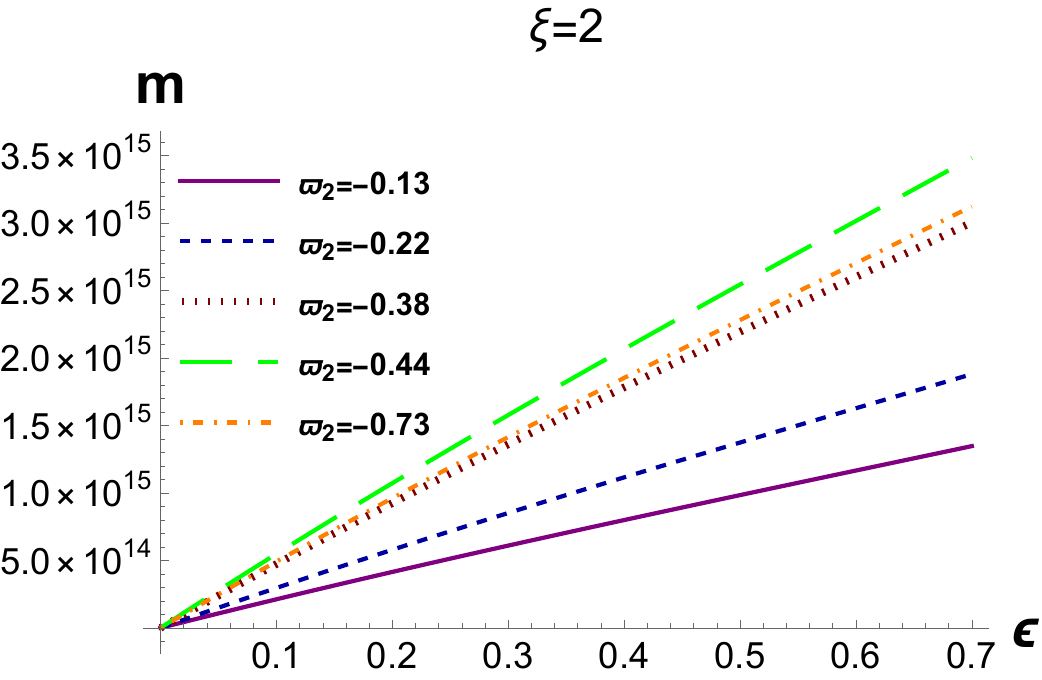}}}
\caption{\label{fig8} Exploring exotic matter content corresponding to specific $\varpi_2$ in the context of a power-law shape function with $r_0=1$.}
\end{figure*}

Integration under the assumption of a power-law shape function leads to \cite{Tripathy},
\begin{eqnarray}
m&=&8\pi\int_{r_0}^{r_0+\epsilon}(\rho+p_r)r^2dr\nonumber\\&\approx&\pi \aleph\sqrt{r_0}\left[8\left(\frac{\epsilon}{r_0}\right)-2\left(\frac{\epsilon}{r_0}\right)^2+\left(\frac{\epsilon}{r_0}\right)^3\right],\label{Eq46}
\end{eqnarray}
where 
\begin{eqnarray}
\aleph=\frac{\sqrt{r_0}}{2(\xi+1)}\left[\frac{(2\varpi_2+\gamma-1)(27+13\varpi_1+2\varpi_2)}{(2\varpi_1\varpi_2+2\varpi_2+\varpi_1^2+3)(8\pi+3\ell_1)-12\ell_1+32\pi\varpi_1}\right].
\end{eqnarray}
Figure~\ref{fig8} illustrates the exotic mass content of wormholes described by a power-law shape function as a function of the parameter $\epsilon$. Specific values of $\varpi_2$, as listed in Table~\ref{tab2}-where the NEC holds-have been considered. For each selected $\varpi_2$, the exotic mass is predominantly governed by the linear term, displaying an almost linear increase with rising $\epsilon$. Moreover, in regions slightly offset from the wormhole throat, the exotic mass content approaches zero. In this case, unlike case 1, where only the value of $m$ for $\varpi_2$ is positive depending on $\xi>-1$, it is positive for all the cases mentioned in Table~\ref{tab2}.
\section{Conclusions}\label{sec7}
In this work, we investigated traversable wormhole solutions in squared-trace extended gravity $\mathfrak{f}(\mathds{R},\mathds{T}) = \mathds{R} + \ell_1 \mathds{T} + \ell_2\mathds{T}^2$ within the framework of Finsler–Randers geometry equipped with the Barthel connection. By considering exponential and power-law shape functions together with anisotropic barotropic equations of state, we demonstrated that mathematically consistent wormhole configurations can be obtained without requiring exotic matter. In particular, the null, weak, strong, and dominant energy conditions were found to be satisfied for suitable parameter ranges, while the corresponding geometries remain regular, horizon-free, and traversable.
\par The main novelty of the present study lies in incorporating the Barthel-type Finslerian structure into squared-trace gravity and showing that the intrinsic anisotropy of Finsler geometry enlarges the parameter space for NEC-satisfying wormhole solutions. We showed that the barotropic equations of state \(p_r=\varpi_1\rho\) and \(p_t=\varpi_2\rho\) admit explicit physically viable regimes, particularly \(\varpi_1\approx 1/3\) and \(\varpi_2\in(-1,-1/3)\), for which all classical energy conditions (NEC, WEC, SEC, and DEC) are simultaneously satisfied. In this configuration, radiation-like behavior in the radial direction generates positive radial pressure, whereas quintessence-like tangential pressure helps maintain the wormhole throat open. Compared with earlier Riemannian \(\mathfrak{f}(\mathds{R},\mathds{T})\) wormhole models, especially Ref. \cite{Tripathy}, the present Finslerian framework permits NEC-satisfying solutions over significantly wider intervals of the tangential equation-of-state parameter, including parameter domains that are inaccessible in the Riemannian limit.
\par We further demonstrated that both exponential and power-law shape functions produce geometrically consistent wormhole configurations satisfying the flare-out condition and exhibiting positive energy density in appropriate parameter domains. The analysis of the exoticity integral showed that the amount of exotic matter can be strongly suppressed or effectively avoided near the throat, indicating that these configurations may be supported by non-exotic matter sources. In addition, the generalized Tolman–Oppenheimer–Volkoff analysis revealed that the hydrostatic and anisotropic forces balance each other, suggesting that the obtained solutions remain in static equilibrium within the present framework.
\par Despite these encouraging results, the present analysis has several limitations. First, the Finslerian parameter \(\xi\) was assumed to be constant in order to obtain analytically tractable solutions; allowing \(\xi(r)\) to vary may lead to richer geometric and physical behavior. Second, the study was restricted to two specific shape functions and simple barotropic equations of state. Moreover, the stability analysis was limited to static equilibrium and did not include dynamical or perturbative stability considerations. These assumptions may restrict the generality of the obtained solutions and should therefore be explored further in future investigations.
\par Overall, the present study shows that the interplay between squared-trace gravity and Finsler–Randers geometry offers a promising framework for constructing traversable wormholes supported by anisotropic matter configurations that satisfy classical energy conditions. Future investigations may extend this framework by considering variable Finslerian functions, more general anisotropic matter distributions, rotating wormhole geometries, or alternative modified gravity theories such as \(\mathfrak{f}(\mathds{Q},\mathds{T})\), \(\mathfrak{f}(\mathds{R},\mathds{G})\), and teleparallel models. It would also be interesting to explore possible observational signatures, including gravitational lensing effects, shadow structures, and quasinormal modes associated with Finslerian wormholes. Such studies may provide deeper insight into the astrophysical relevance of anisotropic wormhole geometries and the broader role of Finsler geometry in gravitational physics. While these solutions are mathematically admissible and satisfy classical energy conditions, a definitive proof of physical reality requires full dynamical stability and causal testing. Future investigations will evaluate these configurations under time-dependent linear perturbations and explicitly track radial subluminal causality bounds.

\section*{Acknowledgements}
The work of Kazuharu Bamba was supported in part by the JSPS KAKENHI Grant Nos. 24KF0100, 25KF0176, Competitive Research Funds for Fukushima University Faculty (25RK011), and a grant-in-aid of academic research of the Yamaguchi Scholarship Foundation.

\end{document}